\documentclass{aa}
\usepackage{graphicx}
\begin{document}
\title{The Local Stellar Population of Nova Regions in the
Large Magellanic Cloud }
\author{Annapurni Subramaniam \& G. C. Anupama}
\institute{Indian Institute of Astrophysics, II Block Koramangala, Bangalore 
560 034, India}
\offprints{Annapurni Subramaniam; e-mail: purni@iiap.ernet.in}

\date{Received / accepted}

\authorrunning{Subramaniam \& Anupama}
\titlerunning{Stellar Population of Novae Regions in the LMC}

\abstract{
This study aims at identifying and understanding the parent population
of novae in the Large Magellanic Cloud (LMC) by studying the local, 
projected, stellar population. The star formation history of the local 
environment around novae is studied based on photometric data of stars 
and star clusters in the nova neighbourhood, available in the 
OGLE II survey and star cluster catalogues. 
The age of the stellar population within a few arcmin around novae regions 
are estimated using isochrone fits to the V vs (V$-$I) colour-magnitude
diagrams. The fraction of stars in various evolutionary states are compared 
using luminosity functions of the main-sequence stars and the red giant 
stars.\\ 
\\
The age, density and luminosity function of the stellar population 
are estimated around 15 novae. The upper limit of the age of the intermediate
stellar population is found to be 4 Gyr in all the regions, excepting
the region around the slow nova LMC 1948. Star formation in these 
regions is found to have started between 4\,--\,2.0 Gyr, with a majority of 
the regions starting star formation at 3.2 Gyr. This star formation event 
lasted until 1.6\,--\,0.8 Gyr. \\
\\
The star 
formation history of the underlying population of both the fast and
moderately fast novae indicate their parent population to be similar
and likely to be in the age range 3.2\,--\,1.0 Gyr. This is in good
agreement with the theoretical age estimates for Galactic cataclysmic 
variables.\\
\\
The region around the slow nova shows a stellar population in the age
range 1\,--\,10 Gyr, with a good fraction older than 4 Gyr. This indicates
that the progenitor might belong to an older population, consistent with 
the idea that the progenitors of slow novae belong to older population.
\\
\keywords{Novae - Large magellanic cloud - star formation - progenitor}
}

\maketitle

\section{Introduction}

Novae belong to the class of cataclysmic variables (CVs), i.e.\ interacting 
close binary systems in which a white dwarf accretes matter from a Roche 
lobe filling companion. Novae undergo outbursts of amplitude $\Delta m=8-15$
in the optical range, reaching magnitudes as high as $M_v \sim -9$. The 
outbursts, which are accompanied by ejection of matter with velocities 
$\ge 300$~km s$^{-1}$, are triggered by a thermonuclear runaway reaction in 
the hydrogen burning shell at the bottom of the accreted layer on the 
surface of the white dwarf. The rate of decline of the nova, which is well 
correlated with the outburst maximum, determines the speed class of the 
nova. The brightness of novae together with the well established relation 
between the absolute magnitude at maximum and the rate of decline make 
these objects valuable as secondary standards for extragalactic distance 
measurements (e.g.\ Capacciolli et al. \cite{c89}; Capaccioli et al.\ 
\cite{c90}; Pritchet \& van den Bergh \cite{pv87}; Della Valle \& Livio 
\cite{dl95}). The use of novae as distance indicators is however complicated
by the lack of understanding of how the properties of novae vary between 
galaxies.

The photometric and spectroscopic development of a nova outburst depend on 
the properties of the accreting white dwarf such as its mass and composition 
(CO or ONeMg type) and also on other factors such as the metallicity in the 
accreted material and the accretion rate (see Kato \cite{k97}, 
Starrfield, Truran \& Sparks \cite{s00}). Some or all these factors may 
depend on the underlying stellar population. A study of the underlying 
population in a nova environment would hence be useful in revealing the 
influence of the various parameters on nova properties.

Attempts to study the progenitors of novae have, in the past, been based on 
their spatial distribution (eg.\ Duerbeck \cite{d84}; Ciardullo et al.\ 
\cite{cd87}; Della Valle \& Duerbeck \cite{dd93}; Della Valle et al.\ 
\cite{dea94}). Population synthesis models of the statistics and properties 
of Galactic CVs and extragalactic novae indicate that the rate of formation 
of CVs, the nova rate and the distribution of novae over speed classes 
depend on the star formation history (SFH) (Yungelson, Livio \& Tutukov 
\cite{ylt97} and references therein). Based on an analysis of the speed 
classes of Galactic and extragalactic novae and their spatial distribution, 
Duerbeck (\cite{d90}) and Della Valle et al (\cite{det92}) suggested the 
presence of two nova populations: fast, bright disk novae and slow, faint 
bulge novae. They also suggested the disk novae originate from more massive 
white dwarfs. Hatano et al (\cite{het97}) and Hatano, Branch \& Fisher 
(\cite{hbf97}) investigated the observability and spatial distribution of 
classical novae in the Galaxy and M31 using a Monte Carlo technique together
with a simple model for the distribution of dust, and found that most novae 
in these galaxies come from the disk population rather than the bulge 
population. Further, the bulge-to-disk nova ratio is similar to the overall 
bulge-to-disk mass ratio of the galaxy. Galaxies such as M33 and the Large 
Magellanic Cloud (LMC) are disk dominated galaxies. Hence most of the nova 
population in these galaxies are expected from the disk population.

A study of the underlying stellar population can be made by understanding the
SFH of the region around novae using clusters and field stars in the region.
Such a study of the environmental effects on the nova population in the 
Galaxy is influenced by the fact that it is not very easy to identify their 
parent population and the age due to (a) the nova distribution is observed 
from within the Galaxy, and (b) mixing of different stellar population in 
the disk of the Galaxy. To study the environmental effects on novae and the 
implications, it is hence ideal to look at an external galaxy where the 
uncertainty in internal distances are eliminated. The LMC is one of the very few 
external galaxies where a large number of novae have been detected and 
studied in detail over the years. The known differences in the abundances 
and evolution between the LMC and the Galaxy might help in understanding 
the influence of these parameters on novae properties.  Further, the LMC is 
believed to have undergone a burst of star formation 3-5 Gyr ago, which 
probably continued to the present day (Butcher \cite{b77}). This event of 
star formation resulted in the majority of the intermediate population seen 
in the disk of the LMC. Thus, a study of the SFH may be able to point to the
age of the parent population of novae in the LMC.

Nova outbursts in the LMC have been recorded since 1926 (Buscombe \& de 
Vaucouleurs \cite{bd55}). A study of the distribution of novae and supernova 
remnants in the Magellanic clouds by van den Bergh (\cite{v88}) indicated 
novae were widely distributed over the face of the LMC with a possible 
clumping slightly to the south of the Bar. van den Bergh also noted that 
there was no concentration of novae within the Bar itself, in contrast with 
the distribution of supernova remnants, and concluded the Bar to be younger 
than the intermediate age population with which novae are associated. 
Subsequent to van den Bergh's study, some of the novae detected in the 
recent past are found to be located in the region of the Bar. 

In the present work, the stellar population 
in the region of novae are studied using star clusters and field stars in 
the nova neighbourhood. Star clusters are used to trace the epochs of 
cluster formation while field stars trace the SFH of the region.
This study, which is the first such attempt, thus aims at identifying 
and understanding the properties of the
parent population of novae by studying the local SFH of the 
novae regions instead of the global SFH of the LMC. Such a study will enable  
inter-comparison of the properties such as the age and metallicity of the
different regions, which may be used to compare and correlate with the 
properties of novae. 

\section{Novae in the LMC and their distribution}

Thirty novae have so far been discovered in the LMC, of which one is a
recurrent nova. Only 19 novae have photometric/spectroscopic data from which
the speed class and other outburst properties are available. Following the
speed classes given by Duerbeck (\cite{d90}), we find that 14 novae belong to
the very fast class ($t_3 < 10$), or fast ($t_3 = 11-25$~days) category. 
Among these, one is super bright (LMC 1991), one is recurrent 
(LMC 1968/LMC 1990\#2) and three are known to be ONeMg novae (LMC 1981, 
LMC 1988\#2, LMC 1990\#1). 3 novae belong to the moderately fast category 
($t_3 = 26-80$~days) and two have been found to be slow novae 
($t_3 > 80$~days). LMC 1995 was discovered as a super soft x-ray source at 
late phases and is analysed to have occurred on a massive ($1.2\, M_\odot$) 
CO white dwarf (Orio \& Greiner \cite{og99}). The recently discovered 
LMC 2002 appears to be a fast nova. LMC 1999 does not appear to be a nova. 
The location, speed class and other known properties of the LMC novae are 
tabulated in Table 1.

\begin{table*}
\caption{List of novae in the Large Magellanic Cloud: coordinates, speed class
and type, references for speed class and type.}
\begin{tabular}{lllllllllll}
\hline
No.&Nova &\multicolumn{3}{c}{RA (2000)}&\multicolumn{3}{c}{Dec(2000)}
&$t_3$& Remarks & Ref.$^{\rm{a}}$\\
&     & h & m & s & $^\circ$ & $^\prime$ & $^{\prime\prime}$ & Days & & \\
\hline
1. &LMC  1926  & 05&14&54.54 &$-66$&48&44.06& 200 & slow nova & 1, 2, 3\\
2. &LMC  1935  & 03&59&15.90& $-67$&46&35.71& 25.1 & fast nova & 1, 2, 3\\
3. &LMC  1936  & 05&07&26.75& $-66$&39&12.08& 31.6 & moderately fast nova & 
1, 2, 3\\
4. &LMC  1937  & 05&57&04.44& $-68$&54&47.92& 19.9 & fast nova & 1, 2, 3\\
5. &LMC  1948  & 05&38&15.38& $-70$&20&26.23& 101.1 & slow nova & 1, 2, 3\\
6. &LMC  1951  & 05&12&51.93& $-69$&58&36.25& 6.26 & v. fast nova & 1, 2, 3\\
7. &LMC  1968  & 05&09&58.28& $-71$&39&51.49& 5.26 & v. fast nova; USco 
type RN & 2, 3, 4, 5\\
8. &LMC  1970\#1 & 05&33&13.25& $-70$&35&04.41&  & poor data & 2\\
9. &LMC  1970\#2 & 05&35&28.90& $-70$&47&14.32& 15.3 & fast nova & 2, 3\\
10. &LMC  1971\#1 & 04&58&23.23& $-68$&05&34.02& 28.3 & moderately fast nova 
& 2, 3\\
11. &LMC  1971\#2 & 05&40&35.22& $-66$&40&35.23&  & poor data & 2\\
12. &LMC  1972  & 05&28&24.66& $-68$&49&42.92&  & poor data & 2\\
13. &LMC  1973  & 05&15&18.58& $-69$&39&46.00&  & poor data & 2\\
14. &LMC  1977\#1 & 06&05&45.50& $-68$&38&12.74&  & poor data & 2\\
15. &LMC  1977\#2 & 05&05&10.87& $-70$&09&01.51& 20.7 & fast nova & 2, 3, 6\\
16. &LMC  1978\#1 & 05&05&52.26& $-65$&53&02.67& 7.8 & v. fast nova & 2, 3\\
17 &LMC  1978\#2 & 05&00&59.65& $-67$&12&44.81& & poor data & 2\\
18. &LMC  1981  & 05&32&09.27& $-70$&22&11.70&  & spectra indicate v. fast; 
ONeMg nova & 2, 7\\
19. &LMC  1987  & 05&23&50.12& $-70$&00&23.50& 5.26 & v. fast nova & 2, 3\\
20. &LMC  1988\#1 & 05&35&29.33& $-70$&21&29.39& 39.2 & moderately fast nova& 
2, 3, 8\\
21. &LMC  1988\#2 & 05&08&01.10& $-68$&37&37.67& 9.7 & v. fast; ONeMg nova &
2, 3, 9 \\
22. &LMC  1990\#1 & 05&23&21.82& $-69$&29&48.48& 7.69 & v. fast; ONeMg nova &
 2, 3, 10\\
23. &LMC  1990\#2 & 05&09&58.28& $-71$&39&51.49& 5.26 & Recurrent nova 1968\\
24. &LMC  1991  & 05&03&44.98& $-70$&18&13.64& $6\pm 1$ & v. fast; super 
bright, low metallicity & 11\\
25. &LMC  1992  & 05&19&19.84& $-68$&54&35.09& 11.23 & fast nova & 3\\
26. &LMC  1995  & 05&26&50.33& $-70$&01&23.08&  & SS X-ray at late phases. 
$1.2\, M_\odot$ CO WD & 12\\
27. &LMC  1997  & 05&04&26.07& $-67$&38&38.00& \\
28. &LMC  1999  & 05&35&32.77& $-69$&29&52.01& & does not appear to be a 
nova\\
29. &LMC  2000  & 05&25&01.60& $-70$&14&17.03& \\
30. &LMC  2002  & 05&36&46.64& $-71$&35&34.4& 23 & fast nova & 13, 14, 15
\\ 
\hline
\multicolumn{11}{l}{$^{\rm{a}}$References: (1) Buscombe \& de Vaucouleurs (1955),
(2) Capaccioli et al.\ (1990) and references therein, 
(3) Della Valle \& Livio}\\ 
\multicolumn{11}{l}{(1995) and references therein, (4) Sekiguchi et al.\ (1990),
(5) Shore et al.\ (1991), (6) Canterna \& Thompson (1981),}\\
\multicolumn{11}{l}{(7) Andrillat \& Dennefeld (1983), (8) Schwarz et al.\ 
(1998), (9) Sekiguchi et al.\ (1989), (10) Vanlandingham et al.\ (1999),}\\
\multicolumn{11}{l}{(11) Schwarz et al.\ (2001), (12) Orio \& Greiner (1999),
(13) Liller (2002), (14) Kilmartin \& Gilmore (2002), (15) Gilmore (2002)}
\end{tabular}
\end{table*}

\begin{figure*}
\centering
\includegraphics[width=17cm]{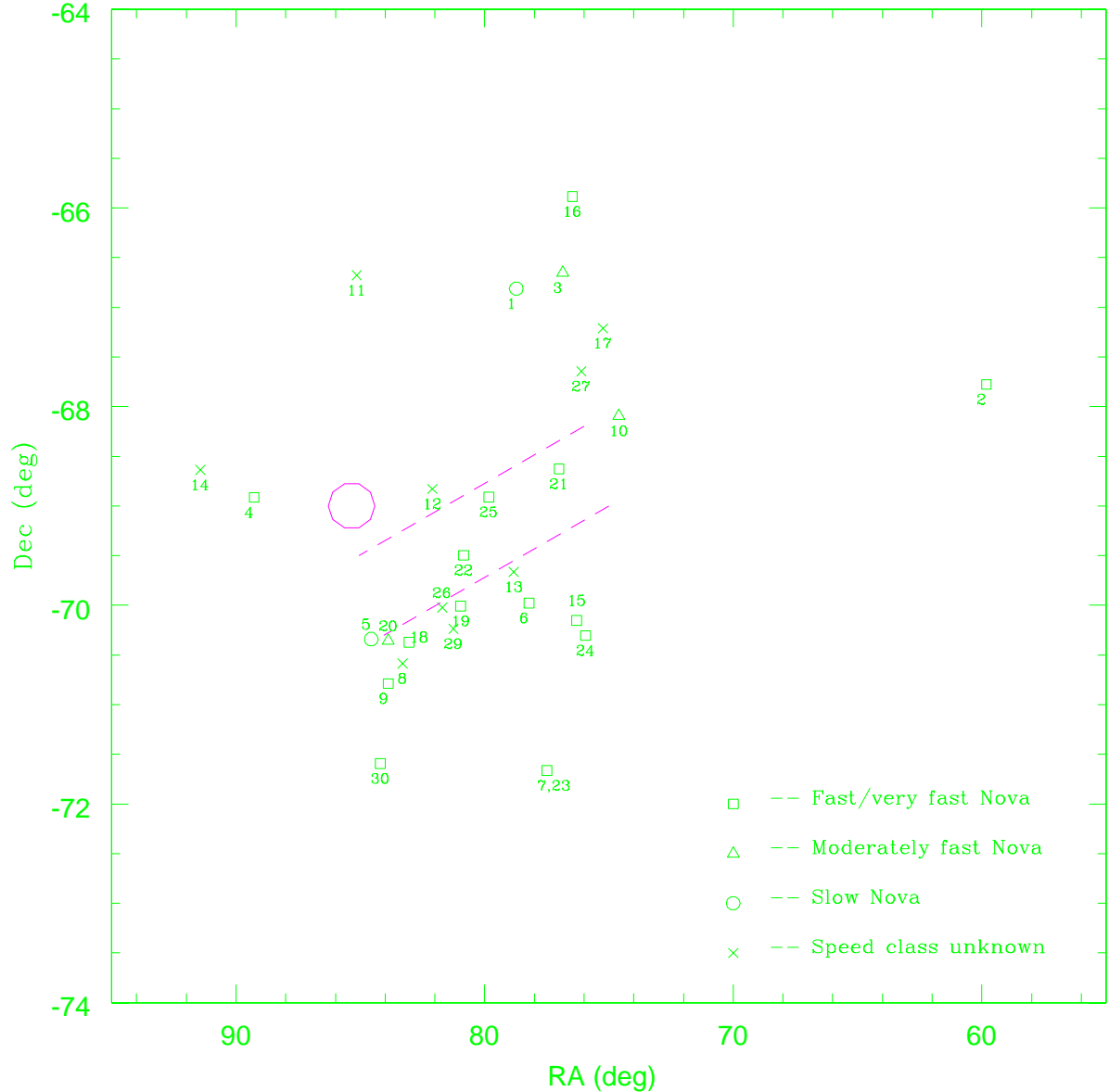}
\caption{Distribution of novae across the face of the Large Magellanic Cloud.
The Bar and 30 Dor regions are also shown.}
\label{figure1}
\end{figure*}
Since the previous study of the distribution of novae in the LMC by van den 
Bergh (\cite{v88}), ten new novae and one recurrent nova have been 
discovered. Figure~\ref{figure1} shows the location of all novae in the LMC,
identified by their serial number in Table 1. From the 
figure, it is seen that novae are widely distributed over the face of the LMC, 
with a lack of novae in the 30 Dor region, and an apparent clumping of novae
towards the south east of the bar.

\section{Novae neighbourhood stellar population}

The ages, density and metallicity of the projected stellar population in the 
neighbourhood of novae are studied based on the star clusters and the field 
star population in that region. As the focus of this study is the 
intermediate age population, the recent ground based surveys of 
the LMC can be used to obtain the data on stellar population. 
Various, extensive surveys have generated a large database of star 
clusters and field stars in the LMC. For the present analyses, the data on 
field stars are obtained from the OGLE II survey (Udalski et 
al.\cite{u2000}). The data from only one survey are used since it is 
essential to have a homogeneity of data for an inter-comparison of the
different regions. All the available catalogues of the LMC star
clusters are used to locate clusters in the neighbourhood of novae. 

\subsection {Data and analyses}
\subsubsection {Field stars}

Field stars within a radius of a few arcmin in the vicinity of novae were
identified from the OGLE II survey. Photometric data were found to be 
available for regions around 15 novae. As more observations were found to 
be available in the I passband, the $V$ vs $(V-I)$ colour-magnitude diagrams
(CMDs) were used in the analyses. All CMDs show a fairly narrow main-sequence
(MS) and a tight red giant branch (RGB) and a red giant clump (RGC). The 
structure of the RGB and the RGC are looked at in greater detail in order to
analyse the properties of the intermediate age stellar population in the 
region, to which the novae presumably belong.

The stellar data are corrected for an assumed reddening of 
$E(V-I)= 0.10$ mag. Although the reddening in the LMC has been shown to be 
varying and clumpy (Udalski et al. \cite{u99}), a single value for reddening
assumed is here, and is found to be satisfactory while fitting the isochrones
to the main-sequence of the CMDs, except in the case of one region, near 
LMC 1948.
Assuming the relation $A_V=3.24 E(B-V)$ and $E(V-I)=1.37 E(B-V)$, the
value of $A_V$ is obtained as 0.355 mag. This value of extinction 
agrees (within errors) with 
that estimated by Dolphin (\cite{d00}). Following Pietrzynski \& Udalski (\cite{pu00}) a 
distance modulus of 18.24 mag for the LMC is assumed here. For this value of 
the distance modulus, one arcmin on the LMC corresponds to 13.4 pc.

The ages of different stellar populations in the vicinity of novae are 
identified by isochrone fits to the CMDs. An isochrone fit to the MS 
identifies the age of the youngest population while the isochrone fits to 
the red giants (RG) identify the ages of the intermediate to old stellar 
population. The isochrones from Bertelli et al. (\cite{bea94}) with 
metallicities, $Z=0.008$ and $Z=0.004$ are used for the age estimation.

The differential luminosity functions (LFs) of the main-sequence (MS) stars 
and the red giants are estimated and are used to compare the relative 
distribution of stars in the CMDs. The LF is estimated by binning the stars 
in magnitudes, such that the width of each bin is 0.2 mag. The stars bluer 
than $(V-I)_0=0.7$ are assumed to be MS stars and stars redder are assumed 
to be red giant stars. In order to estimate accurate LFs, the incompleteness
in the data should be included. For this, incompleteness in the data 
presented by Udalski et al.\ (\cite{u2000}), which is a function of 
brightness and stellar density are used. Three sets of incompleteness data 
are available, depending on the stellar density of the field. By visually 
comparing the stellar density near the location of the novae, the most likely
values of data incompleteness for each region are chosen. A limiting 
magnitude of V=20.7 is chosen for the present study. This limiting magnitude
implies that the stars in the MS are younger than about 1.6 Gyr, while the 
RGB stars are a mixture of both young and old population. This point is to 
be borne in mind in the discussion of the fraction of stars in the MS and 
RGB.

\subsubsection{Star clusters}

The star clusters in the vicinity of novae were identified and their properties
obtained based on the
following catalogs: Pietrzynski et al. (\cite{pea99}: P99)), Bica et al.\
(\cite{bea99}: B99), Bica et al.\ (\cite{bea96}: B96), Pietrzynski and Udalski 
(\cite{pu00}: PU2000). 
B96 present integrated UBV photometry of 624 star clusters and associations 
in the LMC. They estimate the ages of the clusters based on their integrated 
colours and hence classify the clusters into SWB types (Searle, Wilkinson \&
Bagnoulo \cite{swb80}), which is basically an age sequence. This classification can
be used to obtain the approximate age of the clusters. B99 is a revised version of the 
above catalogue and contains about 1808 star clusters for which the position and extent are 
tabulated.  P99 present photometric data of 745 star clusters and their nearby field, of 
which 126 are new findings. PU2000 estimate the ages for 600 star clusters presented in the 
P99 catalogue. The catalogues in B99 and P99 were used to
identify the clusters, while B96 and PU2000 were used to estimate the ages of 
the identified star clusters.

Clusters have been identified within 30~arcmin of the novae using B99.
345 clusters have been identified near 24 novae. Of these, the age estimates 
for 140 clusters could be obtained from PU2000 and B96. 
B96 gives the age of the cluster in terms of groups. As the interest is
in the overall age of the underlying population rather than the ages of the
individual clusters, age groups give a better insight. Therefore, even those
clusters whose exact ages are known are also grouped. The number of clusters 
detected near each nova, the number for which the age is known and the number
of clusters in various age groups are tabulated in Table 2. The clusters are
grouped into the following five age groups: 
\begin{description}
\item (a) clusters with ages $\log \tau < 7.5 $, indicating clusters 
of very young age
\item (b) clusters with ages $7.5 \le \log \tau < 8.0$, indicating clusters
which are relatively young. 
\item (c) clusters with ages $8.0 \le \log \tau < 8.5$,  
equal to 8.0 and less than 8.5, indicating a group of moderately young clusters.
\item (d) clusters with ages $8.5 \le \log \tau < 9.0$, 
\item (e) clusters with ages $\log \tau \ge 9.0$, indicative intermediate population.
\end{description}

\begin{table*}
\caption{LMC Novae: Statistics of the star clusters lying within 30 arcmin.}
\vspace{0.2cm}
\begin{tabular}{lccrrrrr}
\hline
Nova &  No. of clusters & No. of clusters & \multicolumn{5}{c}{Age groups}\\
     &  within 30 arcmin&  with age known & $\le$ 7.5&7.5\,--\,8.0&8.0\,--\,8.5
&8.5\,--\,9.0&$\ge$ 9.0\\
\hline
\multicolumn{8}{l}{ONeMg Novae}\\
LMC 1981 &     15 &12 & 1 & - & 7 & 3 & 1 \\
LMC 1988\#2&      9 & 8 & - & 4 & 1 & 2 & 1 \\
LMC 1990\#1&     17 &13 & - & 3 & 3 & 3 & 4 \\
\\
\multicolumn{8}{l}{Fast Novae}\\
LMC 1937 &      4 & - & - & - & - & - & -\\
LMC 1951 &     24 & - & - & - & - & - & -\\
LMC 1968 &      4 & - & - & - & - & - & -\\
LMC 1970b&     13 & 1 & - & - & 1 & - & - \\
LMC 1977b&      3 & 3 & 1 & - & 2 & - & -\\
LMC 1987 &     24 &19 & - & 2 &11 & 4 & 2 \\
LMC 1992 &      5 & 4 & 1 & - & - & 2 & 1 \\
LMC 2002 &      5 & - & - & - & - & - & - \\
\\
\multicolumn{8}{l}{Moderately fast Novae}\\
LMC 1936 &    14  & 3& -&- & 3& -& -\\
LMC 1971a&     14 & 1 & - & - & 1 & - & - \\
LMC 1988a&     17 &15 & 1 & 1 & 6 & 5 & 2 \\
\\
\multicolumn{8}{l}{Slow Novae}\\
LMC 1926 &      7 & -& -& -& -& -& -\\
LMC 1948 &     37 & 11& -& 1& 7& 3& -\\
\\
\multicolumn{8}{l}{Speed class unknown}\\
LMC 1970a&     29 & 4 & - & - & 2 & 2 & - \\
LMC 1971b&     10 & - & - & - & - & - & -\\
LMC 1972 &     35 & 3 & 3 & - & - & - & - \\
LMC 1973 &     17 & 9 & 2 & 5 & 2 & - & -\\
LMC 1978b&      8 & 5 & - & 1 & 2 & 2 & - \\
LMC 1995 &     27 &22 & - & 4 &11 & 7 & - \\
LMC 1997 &      2 & 2 & - & - & - & 2 & - \\
LMC 2000 &      5 & 5 & - & - & 3 & 1 & 1 \\
\hline
\end{tabular}
\end{table*}

Field stars within a radius of a few arcmin, upto a maximum of 10 arcmin (134 pc) around the
nova are analysed to study the star formation history, while clusters within 30 
arcmin ($\sim$ 400 pc) radius are considered.
A larger radius for the clusters is justified as they are being used to study
the star formation events which took place on relatively larger
scales.

\subsection{Results}

In this section, the location and underlying stellar population around 
individual novae are discussed. Figures~\ref{figure2} -- \ref{figure16} 
present the CMDs and the LFs for the regions discussed. The isochrones with 
their corresponding ages are also shown in the CMDs. The isochrone fits to 
the CMDs were tried using the both the $Z$ values, but $Z=0.008$ isochrones 
were found to fit the CMDs better. Hence unless otherwise specified, the 
isochrones referred to in this section are for $Z=0.008$. The fitting of isochrones 
to the CMDs is based on visual estimation and also 
there is a gap in the age between two consecutive isochrones. These two 
introduce some error in the estimation of the age of the stellar population. This 
error primarily depends on the age of the isochrone considered. The error
introduced in the age estimate is of the order of 10\%.

Table 3 presents the number of field stars considered around each nova 
region for which data are available. Column 2 presents the number of stars
found in the catalogue and column 3 tabulates the total number of stars 
after applying the incompleteness correction. Also the fraction of stars in 
the MS, RGB and RGC, estimated as described in Section 3.1.1 and applying
the incompleteness correction, are tabulated. The CMDs show the number of 
stars presented in the catalogue, whereas the LF plots show the number of 
stars after correcting for the data incompleteness. The statistical error
at each data point in the LF is estimated and is shown as error bar in the 
figure.
\begin{table*}
\caption{LMC Novae: Statistics of field stars in the neighbourhood}
\begin{tabular}{lrrrrrrrr}
\hline
Nova & Observed & Total & MS  & Red & MS  & RG  & Clump  & Clump \\
     & stars &stars & stars & giants & fraction & fraction & stars & fraction \\
\hline
\multicolumn{8}{l}{ONeMg} \\
LMC 1981     & 2472 & 2806.1 &1473.1 &1333.1 &0.525 &0.475 &639.7 &0.480\\
LMC 1988\#2  & 2948 & 3305.6 &2150.1 &1155.4 &0.650 &0.350 &589.6 &0.510\\
LMC 1990\#1  & 5526 & 9160.9 &5283.2 &3877.7 &0.577 &0.423 &1510.1&0.389\\
\\
\multicolumn{8}{l}{Fast}\\
LMC 1977\#2&2317 &2605.4 &1252.6 &1352.7 &0.481 &0.519 & 605.5 &0.448\\
LMC 1987   &4379 &7158.4 &3805.1 &3353.4 &0.532 &0.468 &1429.5 &0.426\\
LMC 1992   &2165 &3550.6 &1941.8 &1608.8 &0.547 &0.453 & 712.9 &0.443\\
\\
\multicolumn{8}{l}{Moderately fast}\\
LMC 1936   &2157& 2497.2 &1606.1 & 891.0 &0.643 &0.357  &466.4 &0.523\\
LMC 1988\#1&3618& 4072.0 &2319.8 &1752.2 &0.570 &0.430  &915.8 &0.523\\
\\
\multicolumn{8}{l}{Slow}\\
LMC 1948   &2762& 3115.5 &1681.4 &1434.1 &0.540 &0.460  &469.1 &0.327\\
\\
\multicolumn{8}{l}{Speed class unknown}\\
LMC 1970\#1&3693& 4176.2 &2034.3 &2141.9 &0.487 &0.513 &1045.8 &0.488\\
LMC 1973   &6739&10977.6 &5984.3 &4993.3 &0.545 &0.455 &2233.0 &0.447\\
LMC 1978\#2&2160& 2494.9 &1540.6 & 954.3 &0.618 &0.382 & 467.5 &0.490\\
LMC 1995   &1797& 2972.2 &1754.0 &1218.3 &0.590 &0.410 & 523.9 &0.430\\
LMC 1997   &3514& 4090.9 &2442.6 &1648.3 &0.597 &0.403 & 869.8 &0.528\\
LMC 2000   &4054& 6741.1 &3771.4 &2969.7 &0.559 &0.441 &1274.6 &0.429\\
\hline
\end{tabular}
\end{table*}

\subsubsection{Fast novae\,--\,ONeMg type}

\paragraph{LMC 1981}
\begin{figure*}
\centering
\includegraphics[width=17cm]{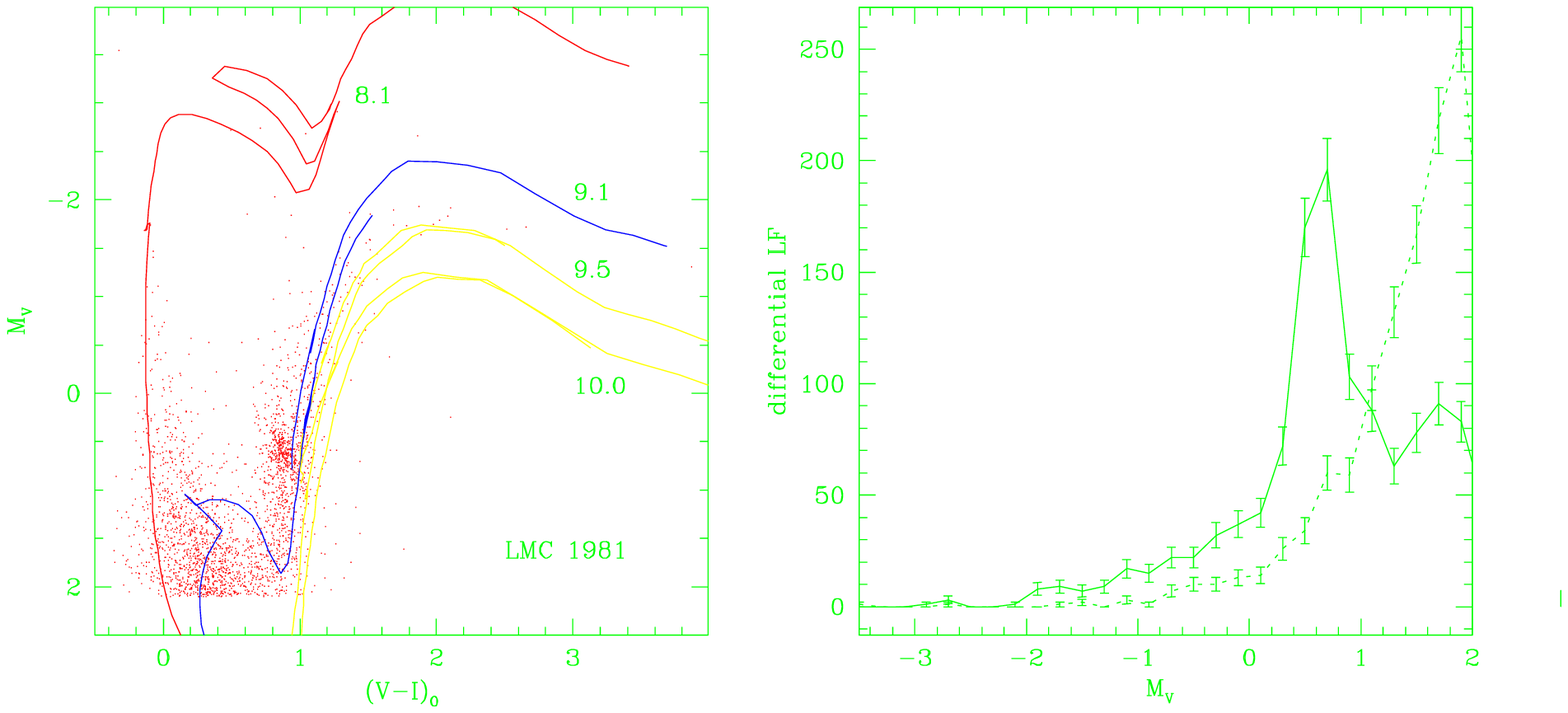}
\caption{Left panel: the CMD of 2472 stars within 3 arcmin from LMC 1981. 
The isochrones fitted to the CMD, with the corresponding value of log(age) indicated, are also 
plotted. Right panel: LF of the MS (dotted line) and the red giants (solid line). The error
bars indicate the statistical error in the data. }
\label{figure2}
\end{figure*}
This nova is located very close to the Bar, slightly south-east of the Bar.
It is one of the 5 novae clustered at this location. Spectroscopy of this
nova during outburst indicate it to be of the ONeMg type (Andrillat \& 
Dennefeld \cite{ad83}).

This nova has 15 star 
clusters within 30 arcmin, of which the ages are known for 12.
58\% of the clusters in this region have ages between 100\,--\,300 Myr, 25\% 
between 300 Myr and 1 Gyr. Only 8\% of clusters are either younger than 30 Myr or
older than 1 Gyr. Hence the bulk of cluster formation has occurred in the 
100-300 Myr range with a tapering towards 1 Gyr.

The field star population within 3 arcmin radius from the nova location 
is studied based on a CMD of 2472 stars. The CMD and the LF are plotted in Figure~\ref{figure2}.
The CMD shows a well defined RGB, RGC and a MS. The RGB does not show much 
scatter. Isochrones corresponding to 10 Gyr, 3.2 Gyr, 1.3 Gyr and 125 Myr are
used to fit the CMD and estimate the ages of the stellar population. These isochrones
are also plotted in the figure.  A few stars belonging to 
the 10 Gyr population are seen in the RGB, as indicated by the 10 Gyr isochrone. 
There seems to be no stellar population between 10 Gyr and 3.2 Gyr. 

Isochrones of ages 3.2 Gyr and 1.3 Gyr fit the RGB and the RGC as well as 
the subgiant branch stars which connect the MS and the RGB at the fainter 
end of the CMD.  The stars towards the left side of the 
1.3 Gyr isochrone at the RGB belong to a younger population. 
The brightest stars in the MS are about 125 Myr old. 

The RGLF plot shows that the red giant branch is populated evenly with a 
clear RGC. The RGC seems to taper to the fainter magnitudes. The MSLF shows 
that the MS does not seem to have very young stars. The upper MS is scantily
populated, with a few stars brighter than $M_V = 0.0$ mag.

The estimated fraction of stars in the CMD reveals that
52.5\% of the stars in this region are in the MS and the rest are in the red giant
phase. Among the red giants, 48.0\% are found in the RGC. 

\paragraph{\bf LMC 1988\#2}
\begin{figure*}
\centering
\includegraphics[width=17cm]{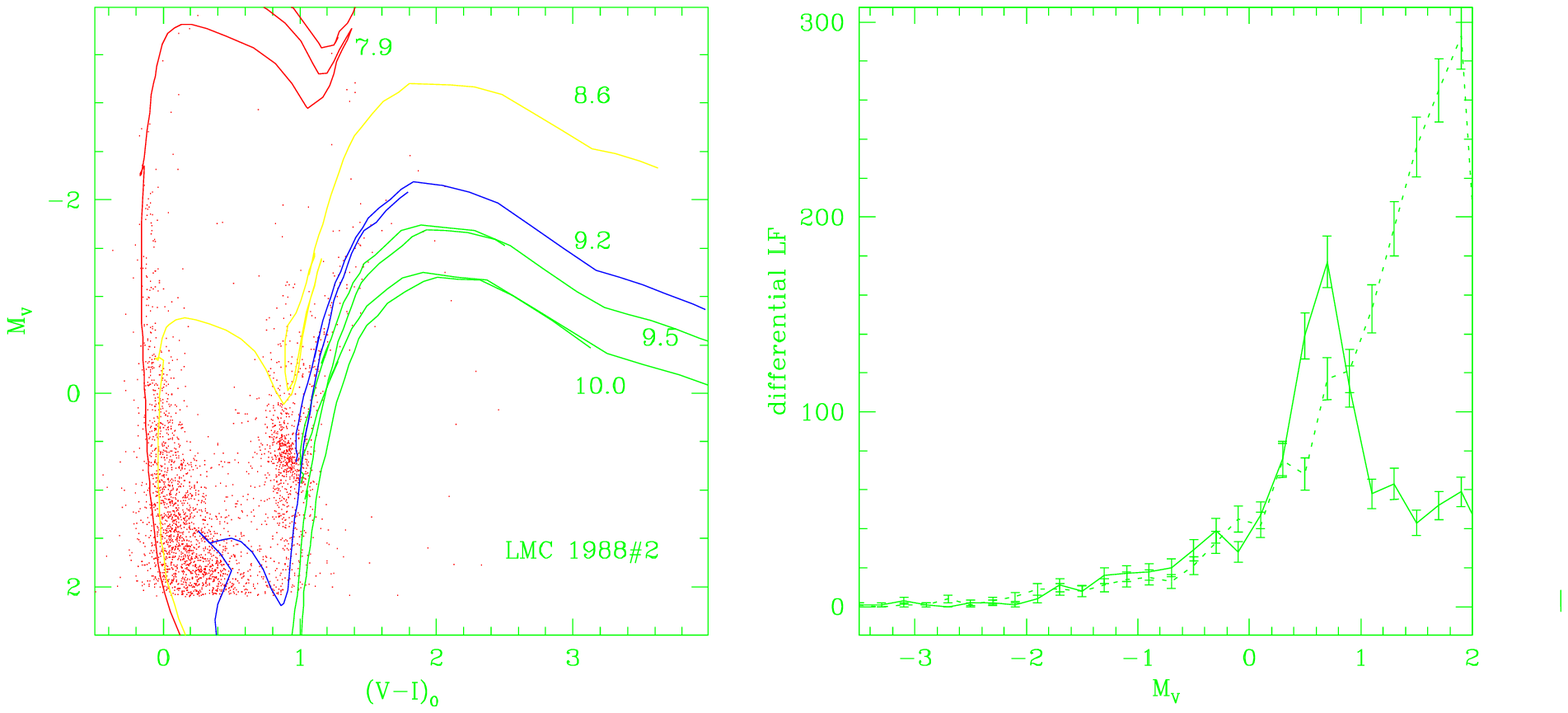}
\caption{Left panel: the CMD of 2948 stars within 4 arcmin from LMC 1988\#2. 
The isochrones fitted to the CMD, with the corresponding value of log(age) indicated, are also 
plotted. Right panel: LF of the MS (dotted line) and the red giants (solid line). The 
error bars indicate the statistical error in the data.}
\label{figure3}
\end{figure*}
This fast nova is located within projected view of the Bar, 
towards its north-western side.
Spectra presented by Sekiguchi et al.\ (\cite{s89}) indicate the nova 
to be an ONeMg type. One supernova remnant of type Ia is found to be located 
close to this nova (Williams et al.\ \cite{w99}). 9 star clusters have been 
identified within 30 arcmin radius, and ages of 8 clusters are known. 50\% 
of the clusters have ages in the range 30\,--\,100 Myr, 12.5\% have ages in 
the range of 100\,--\,300 Myr, 25\% in the range 300 Myr\,--\,1 Gyr and 
12.5\% have ages beyond 1 Gyr. This shows that there has been a constant 
cluster formation, with a rate which doubled during the 300 Myr\,--\,1~Gyr 
period and quadrupled during 30\,--\,100 Myr period. 

2948 field stars located within 4 arcmin around the nova were identified. The
CMD and the LF of the field stars are plotted in Figure~\ref{figure3}. The 
CMD shows a very well populated MS with a fair number of young stars. The 
RGB is well populated with a slightly scattered RGC. The isochrones 
corresponding to the ages 10 Gyr, 3.2 Gyr, 1.6 Gyr, 400 Myr and 79 Myr are 
shown in the CMD. The 3.2 Gyr and 1.6 Gyr isochrones fit the bulk of
the RGB stars. The CMD shows a population of stars directly above the RGC, 
which is like the extension of the RGC. These can be fitted with an isochrone
of age 400 Myr. This is indicative of a star formation event around 400 Myr. 
The brightest stars in the MS are fitted very well with the 79 Myr isochrone, 
indicating that the star formation stopped around 79 Myr ago. 

The RGLF plot shows a well populated RGB with a sharp and strong RGC. The RGB 
is seen to have a recessing peak towards the fainter magnitudes. The MS is well 
populated starting from $M_V = -2.0$ mag and rises gradually beyond $M_V = 0.0$ mag.

This region contains 65.0\% of stars in the MS and
only 35.0\% stars in the red giants. Among the red giants, 51.0\% of stars are 
located in the RGC. 
The relatively high percentage of stars in the MS and the detached MS and RGB at
the fainter end of the CMD show that the stars in the MS are predominantly young. 

\paragraph{\bf LMC 1990\#1} 
\begin{figure*}
\centering
\includegraphics[width=17cm]{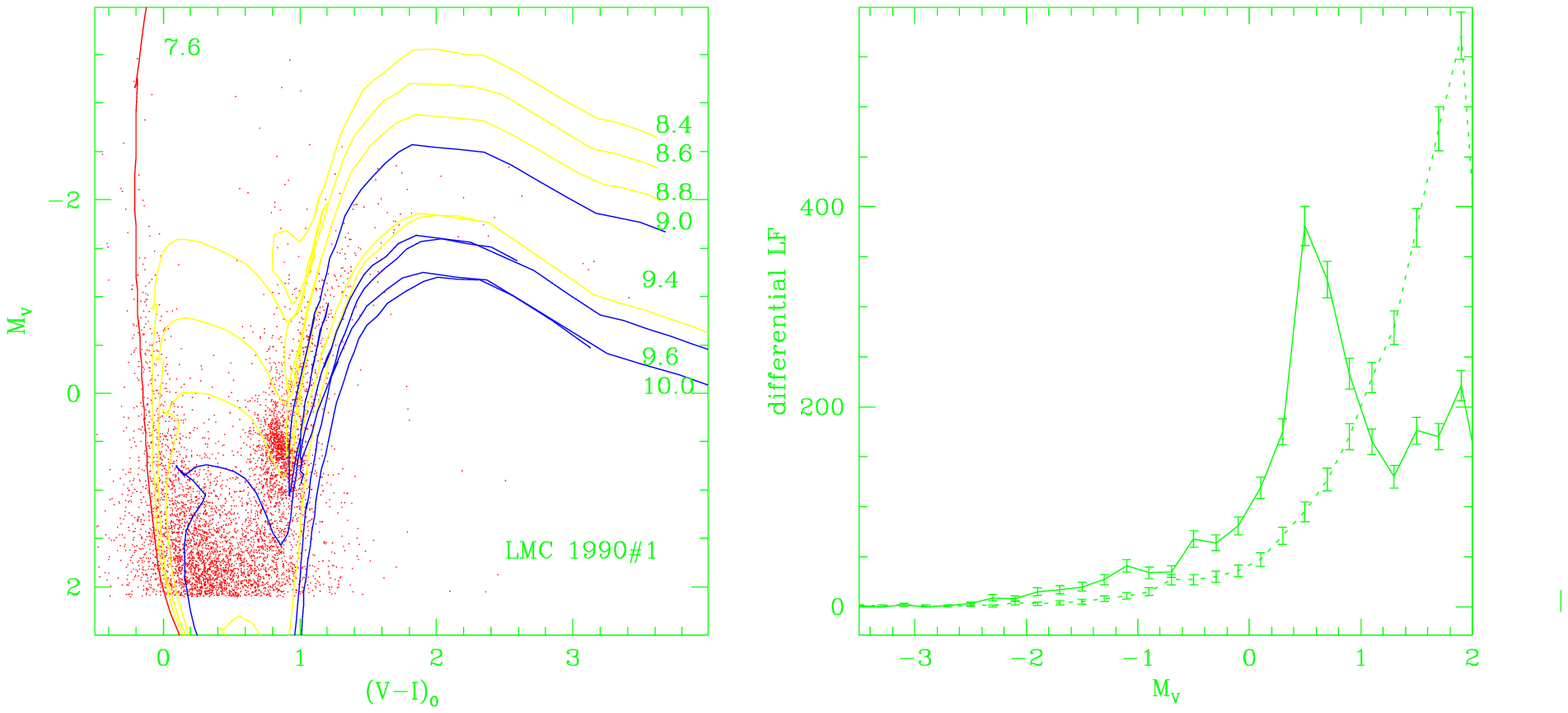}
\caption{Left panel: the CMD of 5526 stars within 2 arcmin from LMC 1990\#1. 
The isochrones fitted to the CMD, with the corresponding value of log(age) indicated, are also 
plotted. Right panel: LF of the MS (dotted line) and the red giants (solid line). The
error bars indicate the statistical error in the data.}
\label{figure4}
\end{figure*}
This ONeMg nova is located in the central region of the projected view of the Bar, 
but closer to the 
southern edge. We find that the stellar density in this region is very high.
Abundance analyses of the outburst spectra (Vanlandingham et al.\
\cite{vea99}) indicate the abundances in the nova material are enhanced in a manner
similar to that seen in the Galactic ONeMg novae. 

Seventeen clusters are found in the vicinity, of which the ages are
known for 13 clusters. Cluster formation in this region appears to be at a 
more or less constant rate until 30 Myr.
31\% of clusters are with ages more than 1 Gyr, and 23\% of clusters
in the lower age ranges. Hence the cluster formation rate was lowered around 
1 Gyr, after an initial higher rate. 

The CMD of 5526 field stars  within a region of 2 arcmin around the nova are
plotted in the left panel of Figure~\ref{figure4}.  The RGB, RGC and the MS 
are well populated. The isochrones corresponding to the ages 10 Gyr, 4 Gyr, 
2.5 Gyr, 1 Gyr, 630 Myr, 400 Myr, 251 Myr and 40 Myr are plotted on the CMD.
A few stars are found to belong to the 10 Gyr population, while most of the 
RGB stars are located between the isochrones of ages 4 Gyr and 250 Myr.
The isochrone fit to the MS of the CMD shows that the brightest MS stars 
are 40 Myr old. Both the field stars and the clusters in this region 
indicate a constant star formation in this region till 30-40 Myr ago.

The RGLF shows that the RGC is less peaked and relatively broader. Also to 
be noticed are the wings on both sides of the RGC peak in the RGLF. The 
MSLF is very smooth. The MS contains 57.7\% and RGB has 42.3\% of stars. 
The shallower peak of the RGC is reflected in the fact that the RGC has 
only 38.9\% of the total number of red giants. All the above support the
suggestion that this region experienced a more or less continuous star 
formation for a longer duration. This result is consistent with that
of Ardeberg et al. (\cite{a97}) who find a continuous star formation history
in the central region of the Bar.

\subsubsection{Fast novae\,--\,others}

\paragraph{LMC 1935}
This fast nova ($t_3=25$~d) is located about $11^\circ$ west of the Bar, in 
a region devoid of any other novae. No clusters are found near this nova 
and also no field star data are available.

\paragraph{LMC 1937}
This nova is located to the east of the 30 Dor region. Photographic magnitude
estimates (Buscombe \& de Vaucouleurs \cite{bd55}) indicate the nova had a 
decline rate of $t_3\sim 20$~days.
There are four star clusters found in the vicinity of this nova. However, the
ages of these clusters are not known. No field star data are
available in the nova region.

\paragraph{LMC 1951}
This extremely fast nova ($t_3=6.3$~d) is located south of the Bar. 
There are 24 star clusters detected near this nova, but the ages of
these clusters are not available. The large number of star clusters near 
this nova implies it is located in a region where cluster formation has
been active. 

\paragraph{LMC 1968/1990\#2}
This is a recurrent nova of USco type (Sekiguchi et al.\ \cite{sk90}; 
Shore et al.\ \cite{sh91} )and is located about 150~arcmin south of the Bar.
Recurrent novae of the U Sco type have been identified as systems 
which could be evolving towards supernovae of type Ia (e.g.\ Starrfield, 
Sparks \& Truran \cite{s85}, Hachisu et al.\ \cite{hea00}).
There are 4 star clusters located near this nova, but their age estimates 
are unavailable. 

\paragraph{LMC 1970\#2 }
This nova is one among the 5 novae clustered at the south-east end of the 
Bar. There is only one star cluster near this nova and the age estimate is 
not available.

\paragraph{LMC 1977\#2 }
\begin{figure*}
\centering
\includegraphics[width=17cm]{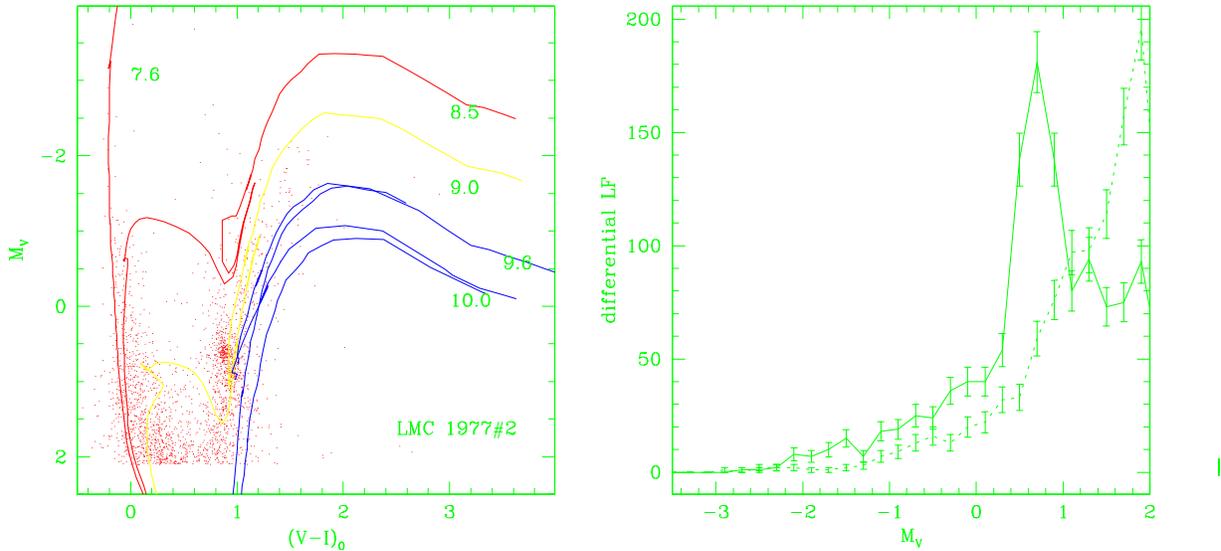}
\caption{Left panel: the CMD of 2317 stars from a nearby region within 
10 arcmin from nova LMC 1977\#2. The isochrones fitted to the CMD, with the 
corresponding value of log(age) indicated, are also plotted. Right panel: LF of 
the MS (dotted line) and the red giants (solid line). The error bars in the figure
indicate the statistical error in the data.}
\label{figure5}
\end{figure*}
This nova is located to the south of the Bar, and also very close to LMC 1991, 
which is also a fast nova. Spectroscopic development during outburst was
similar to Galactic novae (Canterna \& Thompson \cite{ct81}). There are only 
three star clusters near the nova. One 
cluster is younger than 30 Myr and two clusters are in the  range 
100\,--\,300 Myr. The fact that only three clusters are found in the 
neighbourhood indicates that this region may not have experienced high rates 
of cluster formation. 

The CMD of 2317 nearby field stars is plotted in Figure~\ref{figure5}. 
It may be noted here that the field being considered is not centered around 
the nova, but is within 10 arcmin of the nova. Most of the RGB stars are 
populated in the age range 4 Gyr to 1 Gyr, with a few belonging to the
10 Gyr population. The MS and the RGB are well disconnected even at fainter 
magnitudes. 
The MS is well populated and wide upto $M_V = -1.0$ mag, above which, only
a few bright stars are seen. These indicate a 300 Myr old population as
estimated by the isochrone fit. The isochrone fit to the brightest stars in 
the MS indicates these stars are 40 Myr old. 

The RGC is found to have a sub clump towards the fainter end. This feature 
is more clearly seen in the RGLF plot, which is indicated by a second 
peak. 

This region has lesser fraction of stars in the MS, with 48.1\% of stars 
in the MS and the rest 51.9\% in the RGB. In fact, this is one of the two 
regions, where the percentage of stars in MS is less than that in the RGB. 
The RGC contains 44.8\% of the stars in the RGB. It appears that this 
region had a continuous star formation between 4.0\,--\,1 Gyr ago, with 
most of the stars probably having ages closer to the lower age limit, and 
this did not result in the formation of star clusters. Another event of 
star formation took place around 300 Myr, resulting in the formation of the 
two clusters, along with some of the field stars. Then, later, around 
30\,--\,40 Myr, there was another star formation, which managed to form one 
cluster. It is suggested that the lesser fraction of stars in the MS is due 
to a combination of (a) a major fraction of stars being formed in the star 
formation event before $\sim$ 1 Gyr and (b) absence of star formation 
between 1 Gyr and 300 Myr. 

\paragraph{LMC 1978\#1 }
This extremely fast nova ($t_3\sim 8$~d) is located north of the Bar. In fact, 
this is the northern most nova found in the LMC. No star clusters are found near 
this nova.

\paragraph{LMC 1987 }
\begin{figure*}
\centering
\includegraphics[width=17cm]{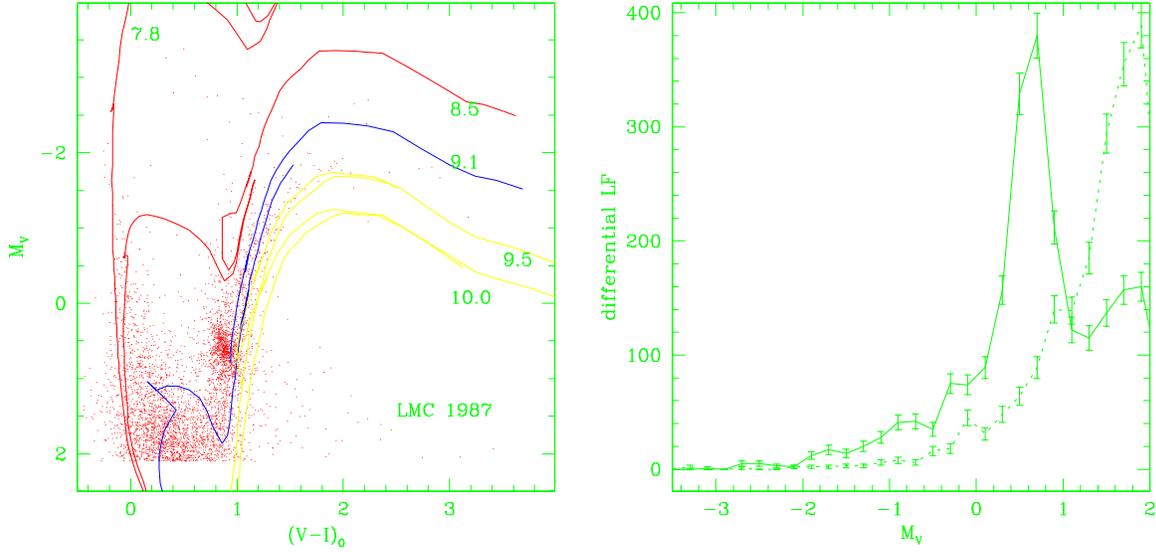}
\caption{Left panel: the CMD of 4379 stars within 2 arcmin from LMC 1987. 
The isochrones fitted to the CMD, with the corresponding value of log(age) 
indicated, are also plotted. Right panel: LF of the MS (dotted line) and the 
red giants (solid line).  The error bars indicate the statistical error in the data.}
\label{figure6}
\end{figure*}
This extremely fast nova is located just below the Bar, in a region that has
produced two more novae; LMC 1995 and LMC 2000. There are 24 star clusters 
in the vicinity and ages are known for 19 of them. This indicates that this 
region had active cluster formation. The maximum number of star clusters 
were formed during the period 100\,--\,300 Myr, when 58\% of the clusters 
were formed. 21\% of clusters were formed during the 300 Myr\,--\,1 Gyr 
period, 10\% formed during the 30\,--\,100 Myr period and the rest during 
the period before 1 Gyr. It appears that this region has been forming 
clusters from the beginning, with an increase in the rate within the last 
1 Gyr, which increased further between 300\,--\,100 Myr. Subsequently the 
rate appears to have reduced and stopped by 30 Myr. 
This in effect implies a continuous cluster formation till 30 Myr. 

The CMD and LF of stars within 2 arcmin radius of the nova are plotted in 
Figure~\ref{figure6}. The isochrone fit reveals that the the stars in the 
RGB have ages in the range 1.3\,--\,3.2 Gyr. Very few stars belonging to 
the age of 10 Gyr is seen in the CMD. The broad MS below $M_V=0.0$ mag 
indicates a continued star formation from about 1.3 Gyr to lesser ages, 
probably till 300 Myr. The brightest part of the MS is found to be 63 Myr 
old. There are a few bright red giants, which appear to be slightly older 
than 63 Myr. The cluster formation and star formation appear to be well
correlated in this region.

The RGLF shows that the RGB is well populated, indicating the presence
of bright red giants. The RGC has a slanting peak. Amongst the total number 
of stars, 53.2\% are in the MS and the rest 46.8\% in the RGB. Further, 
42.6\% of the stars in the RGB are found in the RGC. 

\paragraph{LMC 1991}
This extremely fast nova is located to the south of another fast nova, 
LMC 1977\#2. LMC 1991 was detected as a super bright nova with very low 
metallicity in the ejected material (Della Valle \cite{d91}, Schwarz et al.\
\cite{sw01}). There are 3 very young star clusters within 10 arcmin radius 
from the nova location. 

\paragraph{LMC 1992}
\begin{figure*}
\centering
\includegraphics[width=17cm]{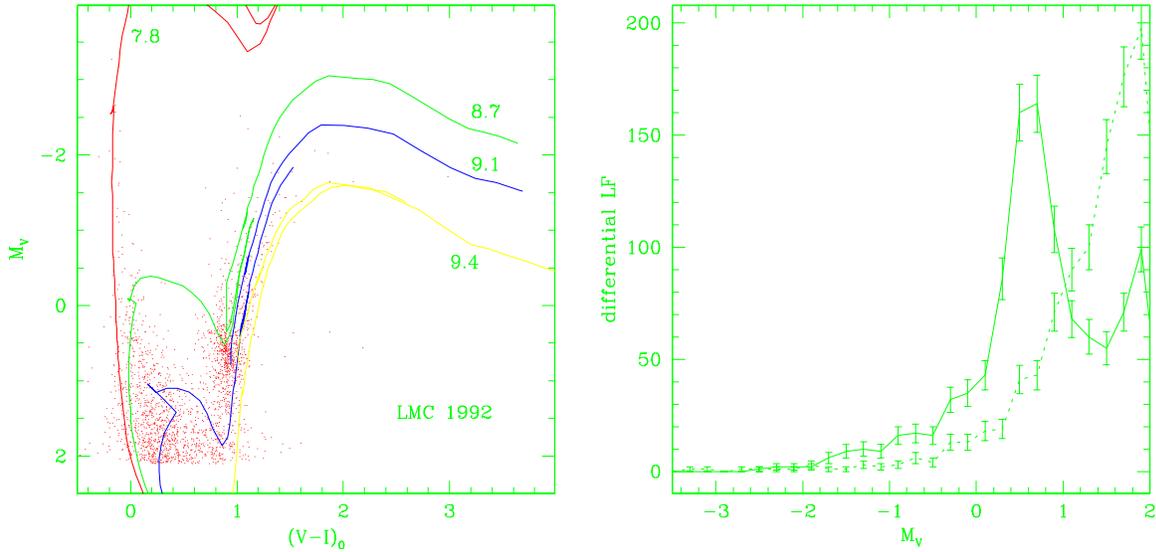}
\caption{Left panel: the CMD of 2165 stars within 4 arcmin from LMC 1992. 
The isochrones fitted to the CMD, with the corresponding value of log(age) 
indicated, are also plotted. Right panel: LF of the MS (dotted line) and 
the red giants (solid line). The error bars indicate the statistical error
in the data.}
\label{figure7}
\end{figure*}
This fast nova is also located in the Bar region.
One type Ia supernova remnant is located close to the nova.
There are five star clusters located near the nova and ages of four have
been estimated. One cluster is with age less than 30 Myr. Two clusters are
in the age range 300 Myr\,--\,1 Gyr and one cluster is older than 1 Gyr. 
The ages and numbers of the clusters indicate that this region did not 
experience much cluster formation. The presence of the young cluster 
indicates that there has been a recent episode of cluster formation. 

The CMD and LF of 2165 stars located within 4 arcmin of the nova, as shown 
in Figure~\ref{figure7}. The CMD indicates an absence of stars belonging
to the old population. The reddest stars in the RGB have an age of
2.5 Gyr, as seen from the isochrone fit. The star formation in this region
seems to have continued till about 500 Myr, as indicated by the corresponding
isochrone fit and the broad MS below this isochrone. The MS is not so well 
populated in the brighter end. The isochrone fit indicates that the 
brightest stars are 63 Myr old. There is no evidence for a strong population between 
the ages 500 Myr and 63 Myr. The star formation history presented by 
the field stars is very similar to the cluster formation history, except 
that no counter part for the youngest star cluster is seen.

The RGLF shows a clear RGC with a flat peak. The RGC profile tapers towards 
the fainter magnitudes. 
The MSLF shows that there are very little stars brighter than $M_v =-1.0$ mag
and the LF rises after that. 54.7\% of stars are located in the MS, and the 
rest 45.3\% are in the RGB. The RGC has 44.3\% of the RGB stars. 
Star formation appears to have started a little later in this region. This 
region appears to be dominated by intermediate age stars and stars aged 
around 500 Myr.

\paragraph{LMC 2002}
This nova was discovered by Liller (\cite{l02}) on 3.1 March 2002, at a
photographic magnitude of $m_{\rm{pg}}=10.5$. A low dispersion spectrum
showed strong H$\alpha$ and H$\beta$ emission lines confirming its 
identification as a nova. Subsequent photometry (Kilmartin \& Gilmore
\cite{kg02}; Gilmore \cite{g02}) indicate a decline rate of
$\sim 0.13$~mag/day ($t_3\sim 23$~days), placing this nova amongst
the fast category.
Five star clusters are found near this nova, but age estimates for
none of these clusters are  available. Also, no field star data are 
available.

\subsubsection{Moderately fast novae}

\paragraph{LMC 1936}
\begin{figure*}
\centering
\includegraphics[width=17cm]{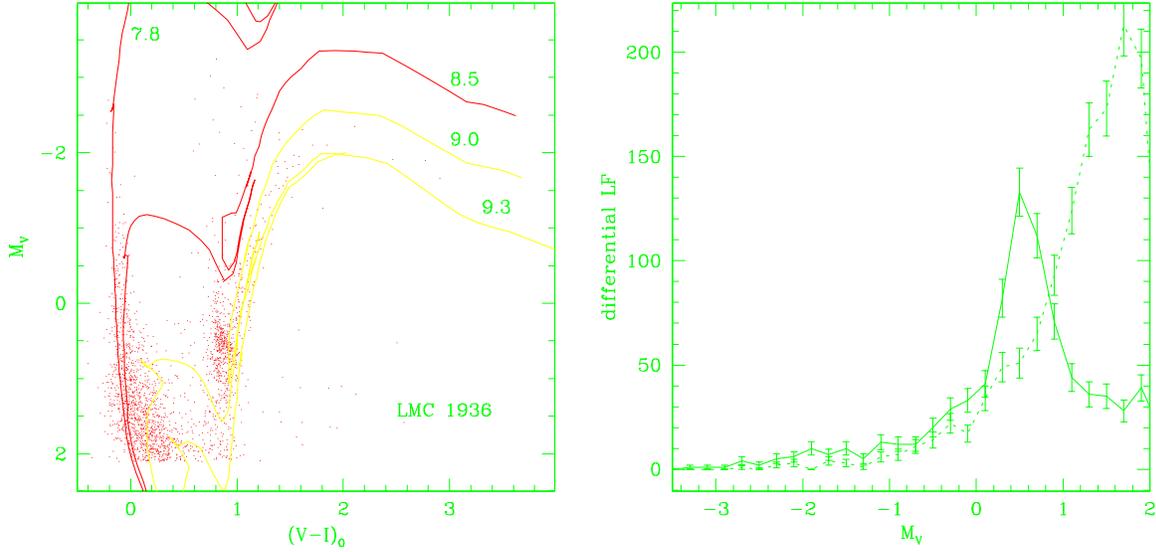}
\caption{Left panel: the CMD of 2157 stars from a nearby region within
10 arcmin from nova LMC 1936. The isochrones fitted to the CMD, with the 
corresponding value of log(age) indicated, are also 
plotted. Right panel: LF of the MS (dotted line) and the red giants 
(solid line). The error bars show the statistical error in the data. }
\label{figure8}
\end{figure*}
This nova is located to the north, about 75~arcmin from the Bar. There are 
14 star clusters found near the nova, of which ages are available only for 3.
All three star clusters have ages in the range 100\,--\,300 Myr. The high 
number of star clusters indicates that cluster 
formation has been quite active in this region. 

The CMD of the 2157 stars belonging to the nearby region, 10 arcmin away 
from the nova, is presented in Figure~\ref{figure8}. The CMD shows that the 
RGB is not very well populated. The 2.0 Gyr, 1 Gyr, 300 Myr and 63 Myr 
isochrone fits to the CMD are shown in the figure. Stars belonging to the 
old population are not found in the CMD. Also the reddest stars in the RGB 
are only 2.0 Gyr old, as indicated by the isochrone fit. If $Z=0.004$ 
isochrones are used, then the reddest stars in the CMD are found to be 
3.2 Gyr old. As we do not have an independent estimate of the metallicity 
and there is no indication of reduced metallicity in this region, the 
$Z=0.008$ isochrones are adopted and the age of the oldest stars is estimated
to be 2.0 Gyr.

The MS and the RGB are well separated at the fainter levels and also show 
that the RGB is very meagerly populated while the MS is well populated. 
The MS shows a sudden decrease in the number of stars brighter than 
$M_V= -1.0$ mag. This indicates another star formation event, and the 
isochrone fit shows that this happened 300 Myr ago. This coincides with the 
formation of the star clusters around this region. The brightest part of 
the MS is estimated to have an age of 63 Myr. 

The RGLF plot shows that the RGC has a ramping region on both sides of the 
peak. The RGC contains 52.3\% of stars in the RGB. 
This region also has a high population of MS stars (64.3\%) when compared 
to the RGB stars, similar to the region surrounding LMC 1988\#2. 

\paragraph{LMC 1971\#1}
This nova is located to the west of the Bar and just outside the Bar. 
There are 14 star clusters found near the nova, of which the age of only one
cluster is known. The age of the cluster falls in the 100\,--\,300 Myr range.
This region has undergone a fair amount of cluster formation, similar to LMC 1936.
No field star data are available for this region.

\paragraph{LMC 1988\#1}
\begin{figure*}
\centering
\includegraphics[width=17cm]{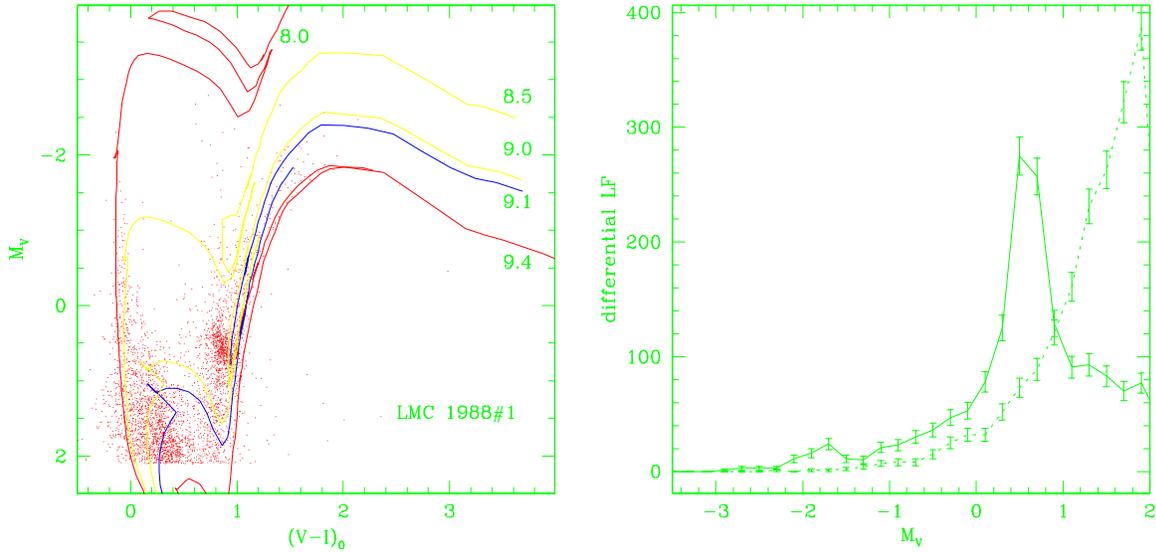}
\caption{Left panel: the CMD of 3618 stars within 2 arcmin from LMC 1988\#1. 
The isochrones fitted to the CMD, with the corresponding value 
of log(age) indicated, are also 
plotted. Right panel: LF of the MS (dotted line) and the red giants 
(solid line). The error bars indicate the statistical error in the data.}
\label{figure9}
\end{figure*}
This nova is located at the south-east edge of the Bar and is one of the 
five clustered novae found in that location. The UV, optical and infrared 
outburst data indicate this nova to be a CO, optically thin dust-forming 
nova (Schwarz et al.\ \cite {sw98}). There are 17 star clusters detected 
near the nova and ages are known for 15 of them. Forty percent of the 
clusters have ages in the range 100\,--\,300 Myr, 33.3\% of clusters
have ages in the range 300 Myr\,--\,1 Gyr, 13.3\% clusters are older then 
1 Gyr. This region thus appears to have experienced a 2.5 times enhancement 
in the formation of clusters, between 300 Myr\,--\,1 Gyr, and increased a 
little more around 300 Myr and continued till 100 Myr. In the last 100 Myr, 
this region has formed 1 cluster older than 30 Myr and one cluster younger 
than 30 Myr. Hence there has been a continuous formation of clusters with 
varying rates. 

The CMD of 3618 field stars within a radius of 2 arcmin from the nova is 
shown in Figure~\ref{figure9}. A broad RGB, a prominent clump and the 
presence of a few bright giants are seen in the evolved part of the CMD. 
Isochrone fit 
to the right most end of the RGB indicates that the oldest stars are 2.5 Gyr 
old. The stars belonging to a population older than 2.5 Gyr is not found. 
At the fainter of the CMD, the MS and the RGB are not well 
separated, but joined together due to the presence of subgiants.
The MS is wide upto $M_V = 0.5$ mag 
corresponding to the turnoff point of the 1 Gyr isochrone. The width of
the MS decreases at brighter magnitudes, indicating a probable decrease in 
the star formation. The isochrone of age 300 Myr shows that the star 
formation probably continued until then. The vertical extension of the RGC, 
as seen in the CMD also indicates that stars younger than 1 Gyr, in the range
500 Myr\,--\,1 Gyr, are present. The brightest part of the CMD is found to 
be 100 Myr old. There is no indication of the presence of stars younger 
than 100 Myr. 

The RGLF indicates some excess stars at the bright end of the RGB, which is 
a combined effect of the old red giants and the bright and young red giants. 
The RGC is seen to have a broad wing on the brighter side, probably due to 
the contribution of stars around 1 Gyr and younger. 
The MSLF shows an evenly populated MS which rises gradually.

The MS has 57.0\% of stars and the rest 43.0\% of stars are in RGB. The RGC 
has 52.3\% of the RGB stars. It is striking to note that the regions around 
both the moderately fast novae have the same fraction of RGB stars in the 
RGC. The fairly high percent of stars in the MS indicates active star 
formation in the last few hundred million years. The star clusters and the 
field stars present a more or less similar star formation history until the 
last 100 Myr. However, while there are no field stars younger than 100 Myr, 
one star cluster younger than 100 Myr is found in this region.

\subsubsection{Slow novae}

\paragraph{LMC 1926}
This nova is located to the north of the Bar and to the east of the 
moderately fast LMC 1936. A supernova remnant of type Ia has been detected 
about half a degree south of the nova. There are 7 star clusters found in 
the vicinity of the nova, but ages of none of the clusters are known. 

\paragraph{LMC 1948}
\begin{figure*}
\centering
\includegraphics[width=17cm]{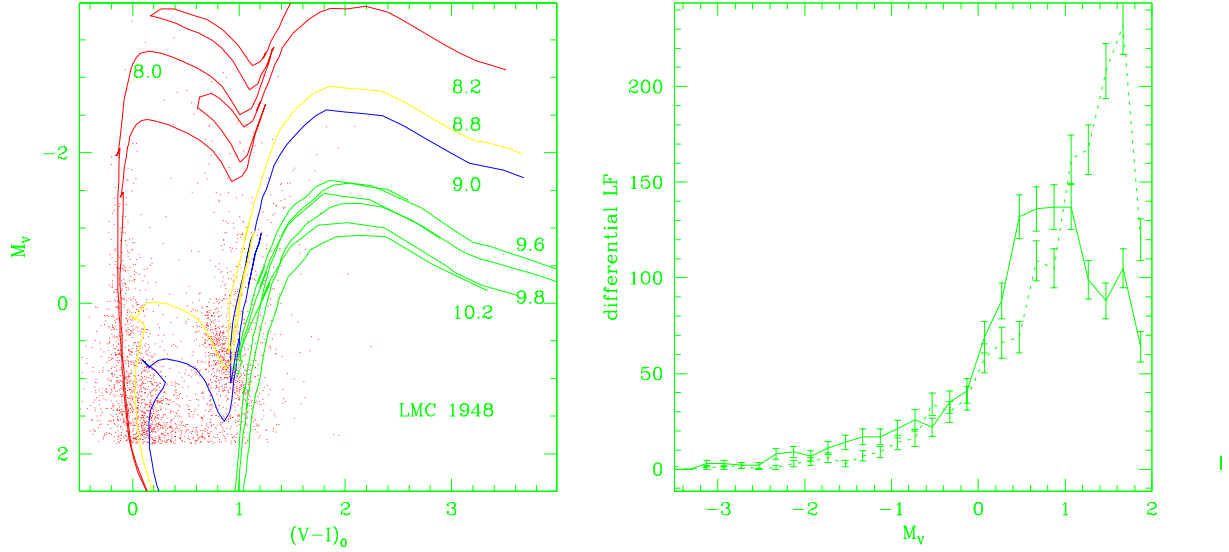}
\caption{Left panel: the CMD of 2762 stars within 2 arcmin from LMC 1948. 
The isochrones fitted to the CMD, with the corresponding value 
of log(age) indicated, are also 
plotted. Right panel: LF of the MS (dotted line) and the red giants 
(solid line). The error bars indicate the statistical error in the data.}
\label{figure10}
\end{figure*}
This is the only slow nova found near the Bar region. This nova also 
belongs to the cluster of novae near the south-east edge of the Bar. 37 
star clusters are found in the vicinity of the nova, of which ages are 
known for 11 of them. None of the clusters are found with age more than 
1 Gyr. 27\% are within the age range 300 Myr\,--\,1~Gyr, 63.6\% are in the 
age range 100\,--\,300 Myr and 9\% have ages less than 100 Myr, but more 
than 30 Myr. Hence the cluster forming activity peaked during 
100\,--\,300 Myr, with clusters forming with lesser efficiencies 
on both sides of this age range. 

The CMD of 2762 field stars within a radius of 2 arcmin from the nova is
shown in Figure~\ref{figure10}. It appears that this region has a higher 
reddening, and a value of $E(V-I) = 0.25$ mag is adopted. The resultant 
CMD indicates that the RGB and the RGC are wide, which could be due to 
differential reddening in the region. However, a close look at the width of 
the MS indicates that it is not too different from the other regions 
studied in this work. It thus appears that the differential reddening is 
not too large and the width of the RGB and the RGC is real. Extremely red 
stars are seen in the RGB, indicative of the presence of very old stars. It 
is found that a 10 Gyr isochrone fits most of the red stars. It is also 
likely that stars older than 10 Gyr could be present in this region. The 
isochrones shown in the CMD are 10 Gyr, 6.3 Gyr, 4 Gyr, 1 Gyr, 630 Myr, 
160 Myr and 100 Myr. It can be seen that the RGB population has the
widest range of ages, from 10 Gyr to 630 Myr. Also, the RGB stars are
evenly populated within this age range, with the 4 Gyr isochrone falling
midway. A simple, visual estimate indicates that approximately half the
RGB population is older than 4 Gyr. At the fainter end of the CMD, the
RGB and the MS are fairly separated with the MS well populated, while
the RGB is scantily populated.  At around 630 Myr, a turn-off near 
the MS is seen, indicating 
a star formation event, which could have resulted in the formation of a 
few star clusters. Star formation might have continued further. A well 
populated MS as seen in the CMD also indicates the same. The brightest 
stars in the CMD are found to be 100 Myr old. The bright red giants are 
found to be 160 Myr old. 

The RGLF profile in this region is quite interesting. It shows a very flat 
RGC, quite unlike the RGC seen in the other regions. The prominent peak of 
the RGC, which is seen in the other regions is missing, indicating a marked 
difference in the star formation history, for ages beyond 1 Gyr. The RGLF 
rises initially showing the presence of a good number of bright red giants. 
Another important point to be noted is that at the fainter end of the RGC 
profile, which is a contribution from older stars, the fraction of stars is 
relatively higher.

The MS has 54.0\% of the stars and RGB has 46.0\% of the total stars
found in this region. Also, the RGC has only 32.7\% of the stars in the RGB. 
This value is very small compared to the other regions. This indicates that 
the star formation occurred in a continuous fashion for a duration much 
longer than in other regions. 

A point of particular interest to note is the significant presence of an old 
star population, up to an age of about 10 Gyr. This population is absent in 
the regions of other novae in the Bar region. This point is of particular 
significance as LMC 1948 is the only slow nova in this region. All other 
novae detected near the Bar region are either fast or moderately fast.

\subsubsection{Others}

The stellar population around the locations of novae for which the speed 
class is not available are discussed in this section.

\paragraph{LMC 1970\#1}
\begin{figure*}
\centering
\includegraphics[width=17cm]{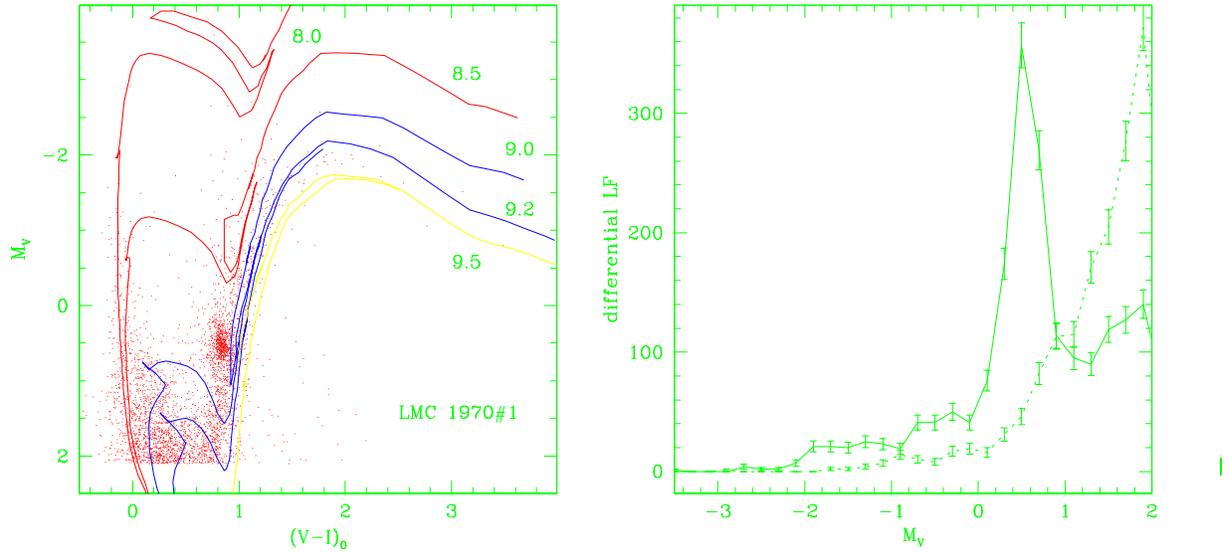}
\caption{Left panel: the CMD of 3693 stars within 5 arcmin from LMC 1970\#1. 
The isochrones fitted to the CMD, with the corresponding 
value of log(age) indicated, are also 
plotted. Right panel: LF of the MS (dotted line) and the red giants 
(solid line). The error bars indicate the statistical error in the data. }
\label{figure11}
\end{figure*}
This nova is located near the south-east edge of the Bar. This is one among 
the five novae clustered in this region. 29 star clusters are found around 
this nova, of which ages are known for only 4 novae. Two clusters are in 
the age range 100\,--\,300 Myr and the other two in the 300\,--\,1 Gyr range.
The large number of clusters indicate that there 
has been a significant amount of cluster formation in this region.

The CMD of 3693 stars within a radius of 5 arcmin of the nova is shown in 
Figure~\ref{figure11}. Isochrones of ages 3.2 Gyr, 1.6 Gyr and 1 Gyr are 
fitted to the RGB. There is no clear signature of stellar population older 
than 3.2 Gyr. This star formation continued till about 1 Gyr, as implied by 
the reduction in the width of the MS above the 1 Gyr isochrone. The wider 
part of the MS at $M_V=-1.0$ and the bright red giants indicate that the 
star formation continued, probably till 315 Myr. The brightest stars in the 
CMD are found to be 100 Myr old.  

The RGC is a narrow and tall peak, indicative of a strong
star formation event. The MSLF shows that the profile 
rises slowly. This region is one of the two regions, other than the region around
LMC 1977\#2, with lesser fraction of stars in the MS, than in the RGB. This region
has 48.7\% stars in the MS.  The RGC has 48.8\% of the stars in the RGB. 
 
\paragraph{LMC 1971\#2}
This nova is located to the north of the Bar and 30 Dor. This is an isolated 
nova, away from the region where most novae are detected. There are 10 star 
clusters near the nova and the ages are not known for any cluster. 
No field star data are available.

\paragraph{LMC  1972}
This nova is located very close to the projected location of the Bar, 
towards the north. 
There are 35 star clusters around the nova. Ages of three clusters are known and
they are all younger than 30 Myr. The large number of clusters and the presence of
very young clusters indicate that the cluster formation has been very vigorous in 
this region. Field star data are not available for this region.

\paragraph{LMC 1973}
\begin{figure*}
\centering
\includegraphics[width=17cm]{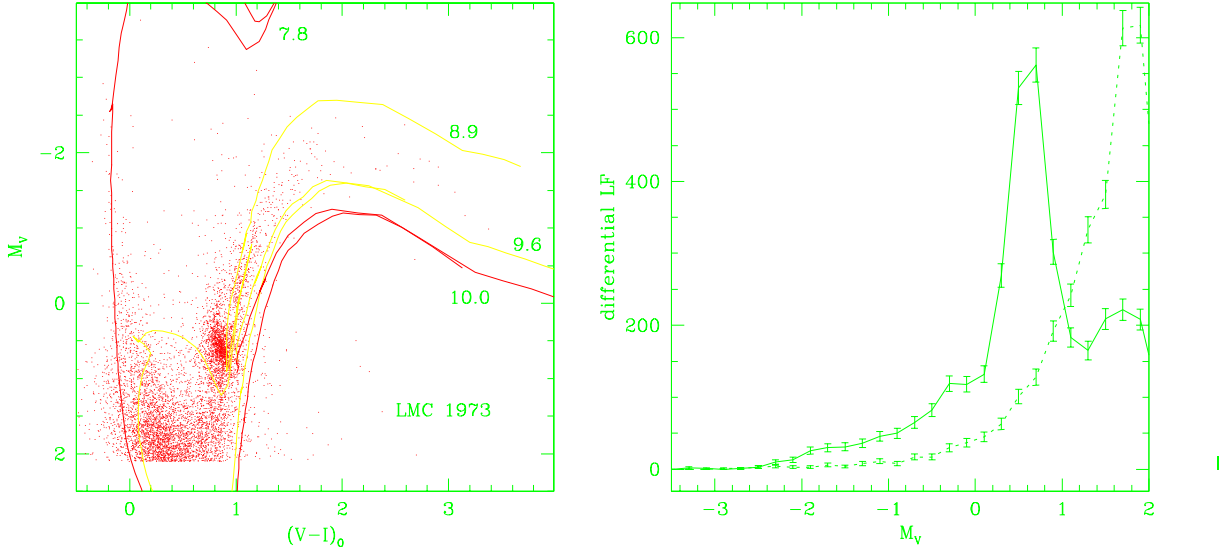}
\caption{Left panel: the CMD of 6739 stars within 3 arcmin from LMC 1973. 
The isochrones fitted to the CMD, with the corresponding 
value of log(age) indicated, are also 
plotted. Right panel: LF of the MS (dotted line) and the red giants 
(solid line). The error bars indicate the statistical error in the data.}
\label{figure12}
\end{figure*}
This nova is located very close to the Bar region, on its southern side.
There are 17 star clusters near this nova, of which ages are known 
for 9 star clusters. 55.6\% of the clusters have ages 
in the range 30\,--\,100 Myr. 22.2\% of star clusters have ages less than
30 Myr and 22.2\% have ages between 100\,--\,300 Myr. From the cluster 
data it is seen that this region had an active cluster formation event in the
recent past, that is 30\,--\,100 Myr, with a little subsided activity before
and after that. As there are 17 clusters, it is more or less clear that this
region has had strong cluster formation episodes. 

The CMD of 6739 field stars which lie within a radius of 3 arcmin from the 
nova is shown in Figure~\ref{figure12}. The isochrones of ages 4 Gyr, 
800 Myr and 63 Myr are shown in the CMD. The merged MS and RGB at the 
fainter end of the CMD shows that the star formation was more or less 
continuous between 4 Gyr and 800 Myr. The CMD indicates this region has a small 
fraction of an old population of around 10 Gyr. Subsequently, perhaps 
around 4 Gyr, star formation started vigorously and continued till 800 Myr, 
following which, until 63 Myr back, star formation occurred in a more 
subdued manner.

The RGLF shows a smooth curve for the RGB and a not so narrow peak for the 
RGC. The smoothly rising RGB profile indicates that star formation has been
more or less continuous. The MS is also a gradually rising smooth profile. 
This region has 54.5\% of stars in the MS and RGC has 44.7\% of the RGB 
stars.

\paragraph{LMC 1977\#1}
This nova is located about $5^\circ$ north-east of the Bar and is the 
eastern most nova detected in the LMC. The fast nova LMC 1937 is located 
to its south-west. No clusters are detected near this nova. Also, no field 
star data are available for this region.

\paragraph{LMC 1978\#2}
\begin{figure*}
\centering
\includegraphics[width=17cm]{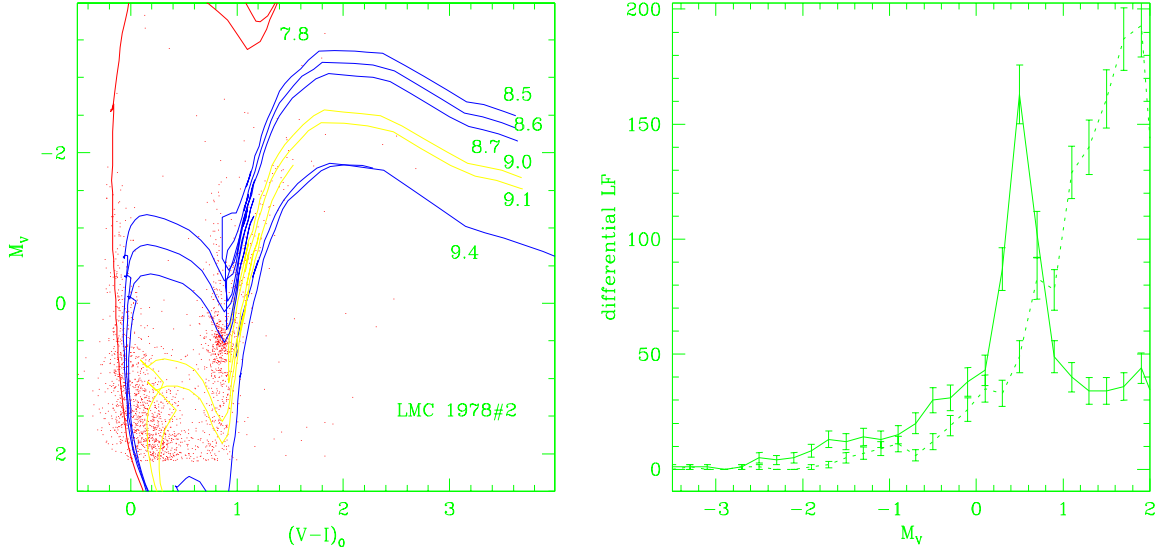}
\caption{Left panel: the CMD of 2160 stars from a nearby region within
5 arcmin from nova LMC 1978\#2. The isochrones fitted to the CMD, with the 
corresponding value of log(age) indicated, are also 
plotted. Right panel: LF of the MS (dotted line) and the red giants 
(solid line). The error bars indicate the statistical error in the data. }
\label{figure13}
\end{figure*}
This nova is located at the north of the western edge of the Bar. There are
8 star clusters found near this nova and ages are known for 5 of them.
The age ranges 100\,--\,300 Myr and 300 Myr\,--\, 1 Gyr are found to 
have two clusters each.  One cluster is found within the age range 
30\,--\,100 Myr.
This shows that there has been continuous formation of clusters, probably
in a subdued manner. 

The CMD and LF of a region close to the nova, within 5 arcmin is shown 
in Figure~\ref{figure13}. The CMD is characterised by a very 
a prominent RGC and a vertical extension of the RGC.
Also, the MS and the RGB are well separated, except for some identifiable
subgiant branches. The isochrone fits reveal that the oldest stars in this
region are 2.5 Gyr and are scantily populated. The wide MS near to $M_V = 1.0$ mag
probably belongs to 1.3 Gyr old population. This isochrone also fits most of the
red giants. The 1.0 Gyr isochrone fits the RGC quite well.
Isochrones with $\log(\tau)=8.5,\, 8.6,\, 8.7$ are seen to fit  
the vertical extension of the RGC. Some 
bright stars are seen at the top of the MS, whose ages are estimated as 63 Myr.

The RGLF shows that the RGB presents a gradually rising curve with a strong
and narrowly peaked RGC. The RGB has a good number of bright stars, which are the
vertical extension of the RGC.

This region has a significantly high fraction of stars (61.8\%) in the MS. 
The RGC has 49.0\% of stars found in the RGB. 
The star formation which started at 2.5 Gyr, continued till 1.0 Gyr.
Another star formation event has occurred at about 400 Myr, which produced 
four star clusters.

\paragraph{LMC 1995}
\begin{figure*}
\centering
\includegraphics[width=17cm]{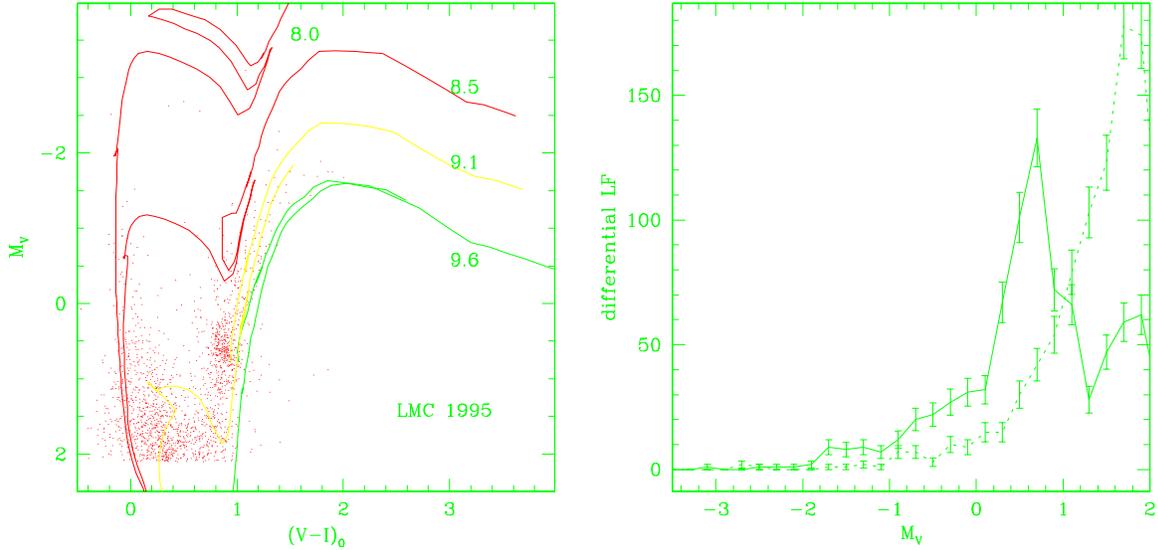}
\caption{Left panel: the CMD of 1797 stars within 3 arcmin from LMC 1995. 
The isochrones fitted to the CMD, with the corresponding value of log(age) 
indicated, are also plotted. Right panel: LF of the MS (dotted line) and the red giants 
(solid line). The error bars indicate the statistical error in the data.}
\label{figure14}
\end{figure*}
This nova is located at the southern edge of the Bar region. This nova has been
detected as a super-soft X-ray source with a $1.2\, M_\odot$ CO white 
dwarf (Orio \& Greiner \cite{og99}). 27 star clusters 
are found in the vicinity of the nova, of which the ages are known for 22. 
18\% of the clusters have ages within 30\,--\,100 Myr, 50\%  have ages 
between 100\,--\,300 Myr and 32\% have ages between 300 Myr\,--\,1 Gyr. 
This indicates that there was a burst of cluster formation some time 
during 100\,--\,300 Myr with a tapering cluster formation before and after 
this burst. 

The CMD  of 1797 stars within a radius of 3 arcmin of the nova is shown in 
Figure~\ref{figure14}. The isochrone fits reveal that the oldest traceable 
population is about 4.0 Gyr. The star formation then continued to about 
1.3 Gyr. This is indicated by the fit of the isochrone to the subgiant 
branch location and the red giants of the RGB. A few stars can be seen to 
the left of the 1.3 Gyr isochrone at the RGB, which indicates that the star 
formation continued until more recent times. There is a clear signature of 
the 300 Myr old population as indicated by the isochrone. The brightest 
stars in the MS are 100 Myr old. 

The RGLF shows that the RGB has a bumpy profile. The RGC profile shows that 
there are two clumps, the brighter clump belonging to a younger population 
and the fainter one to a population older than that. The MSLF also shows a 
bumpy profile. This region has 59.0\% of stars in the MS and the RGC has 
43.0\% of stars in the RGB. 

\paragraph{LMC 1997}
\begin{figure*}
\centering
\includegraphics[width=17cm]{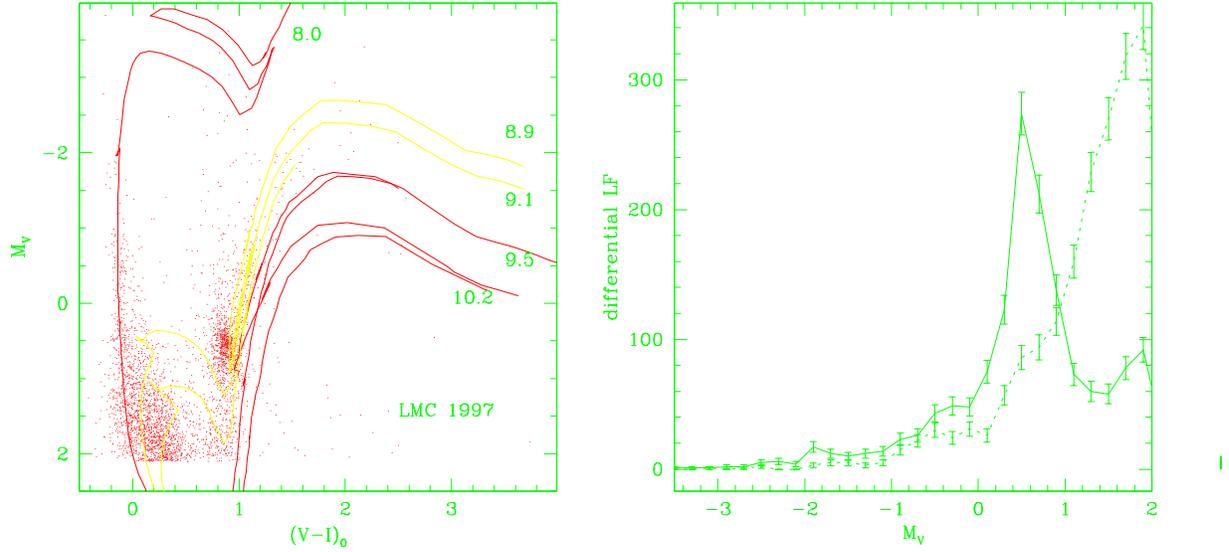}
\caption{Left panel: the CMD of 3514 stars from a nearby region within 
8 arcmin from the nova LMC 1997. The isochrones fitted to the CMD, with the 
corresponding value of log(age) indicated, are also 
plotted. Right panel: LF of the MS (dotted line) and the red giants 
(solid line). The error bars indicate the statistical error in the data.}
\label{figure15}
\end{figure*}
This nova is located to the north of the Bar and slightly away from it.
There are two star clusters found near this nova. Both clusters have an age
between 300 Myr\,--\,1 Gyr. The fact that there are only two star clusters 
in the vicinity indicates that the cluster formation has been very poor. 

The CMD of 3514 stars from a nearby field within 8 arcmin of the nova
is shown in Figure~\ref{figure15}. 
The isochrone fit to the right most part of the RGB shows that some stars 
belonging to about 15 Gyr is present. The fit of the isochrone of age 
$\log\tau= 10.2$ is shown in the CMD. The RGB is not populated with stars 
in the age range 15 and 3.2 Gyr. Subsequent to 3.2 Gyr, star formation 
seems to have been continuous for a long time, as indicated by the broad 
RGB, broad MS, subgiant branches and the broad and prominent RGC. The 
isochrones shown in the figure are for ages 15 Gyr, 3.2 Gyr, 1.3 Gyr and 
800 Myr. The presence of stars on to the left of the 800 Myr isochrone at 
the RGB is indicative of continued star formation to more recent times. The 
well populated MS also supports this. The brightest part of the CMD has 
stars of age 100 Myr. The most luminous red giants also seem to belong to 
this population. 

The RGLF shows that the brighter end of the RGB is very well populated. The RGC 
profile shows a tapering towards brighter magnitudes, and the profile itself is 
fairly broad and well peaked. The MSLF shows a bumpy profile.

This region has 59.7\% of stars in the MS and the rest 40.3\% stars in the RGB. 
The RGC has 52.8\% of the stars in the RGB. This high value 
indicates that star formation was high in the 3.2 Gyr\,--\,800 Myr period. The
star formation appears to have continued until 100 Myr, albeit a lesser vigour.
This region has some trace of the very old population of the LMC. 

\paragraph{LMC 2000}
\begin{figure*}
\centering
\includegraphics[width=17cm]{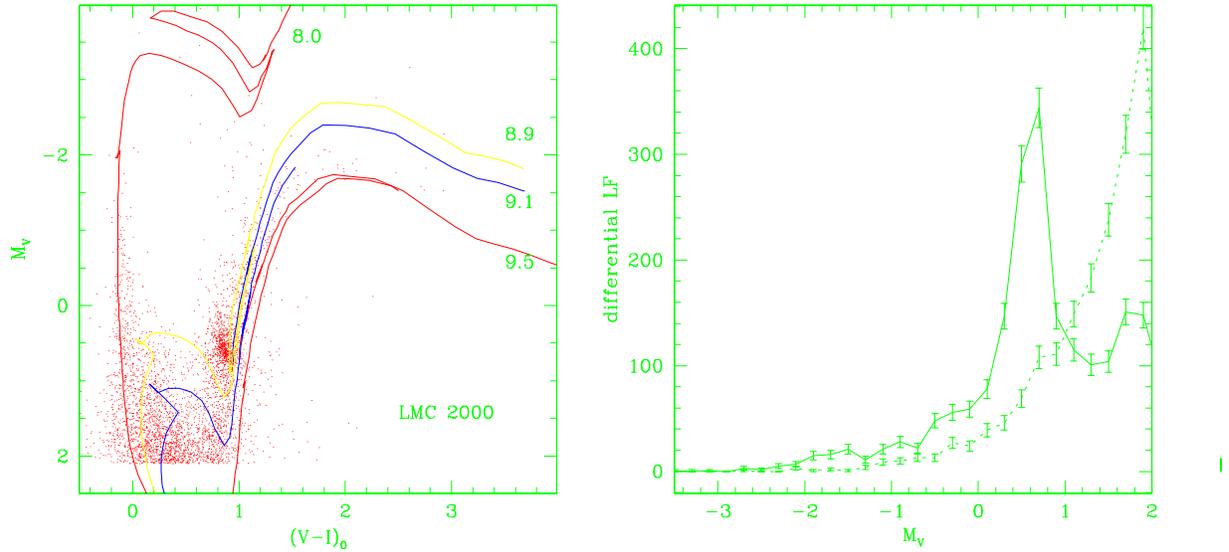}
\caption{Left panel: the CMD of 4054 stars within 3 arcmin from LMC 2000. 
The isochrones fitted to the CMD, with the corresponding value of log(age) 
indicated, are also plotted. Right panel: LF of the MS (dotted line) and the 
red giants (solid line). The error bars indicate the statistical error in the data.}
\label{figure16}
\end{figure*}
This nova is located a to the south of the Bar. There are 5 star clusters 
found near this nova. All the five have age estimates, with one cluster 
each in the age ranges 300 Myr\,--\,1 Gyr; 1 Gyr and beyond respectively. 
Three clusters have ages within 100\,--\,300 Myr. 

The CMD of 4054 stars within a radius of 3 arcmin from the nova is
show in figure~\ref{figure16}. The CMD shows a well populated RGB, MS and a prominent RGC.
The CMD does not show any trace of an old population. The star formation 
probably started at about 3.2 Gyr, as seen by the isochrone fit to the CMD. It 
continued till 800 Myr, as seen by the isochrones of ages 1.3 Gyr and 800 Myr. 
The star formation then stopped or continued at very low rates till 100 Myr ago. 
The MS between the 800 Myr and 100 Myr turn-offs is only scantily populated 
indicating a low star formation rate. 

The RGLF shows a slowly rising profile, with a relatively sharp RGC profile.
MSLF is seen to be a smooth profile. The MS contains 
55.9\% of total number of stars in the region and the rest are in the RGB. 
The RGC contains 42.9\% of stars in the RGB.

\section{Discussion}
\subsection{Distribution and population of novae in the LMC}
A study of the distribution of novae by van den Bergh (\cite{v88}) showed 
that the nova population followed the old disk population. Further,
van den Bergh did not find any nova in the Bar region and concluded the Bar in the 
LMC to be a recent phenomena. Subsequent to van den Bergh's study, 10 novae 
have been detected. The projected locations show that, among these ten,
four are in the Bar and four are very close 
to the Bar. This shows that the population responsible for producing novae 
may also be present in the Bar region.

The LMC seems to have a high population of fast novae, 72.2\% of the novae 
with known speed classes are fast novae. The slow novae are only 11.1\%. 
Duerbeck (\cite{d90}), Della Valle et al.\ (\cite{det92}) and Yungelson, 
Livio \& Tutukov (\cite{ylt97}) have shown the existence of two nova 
populations in the Galaxy and M31: fast, bright disk novae and slow, faint 
bulge novae. The disk novae probably belong to a relatively younger parent 
population having higher white dwarf masses. The larger population of fast 
novae implies a predominantly disk population in the LMC. This is consistent 
with the fact that in the LMC the old halo population is a minority 
population, whereas the intermediate age population of a few Gyr is the 
dominant population. 

The locations covered by the fast and the moderately fast novae are indicative of 
the extension of the LMC disk. It is of interest to note that both the identified
slow novae are located within the regions of fast novae. The recurrent nova is 
also seen to be located not very far from the location of other novae.

\subsection {Star formation history}
The star formation history of nearby regions around 15 novae are studied 
here. The results derived in the previous section are presented in 
Table 4. 

\begin{table*}
\caption{LMC Novae: Age estimates of the neighbourhood stellar population}
\begin{tabular}{lcccc}
\hline
Nova & old stellar & intermediate & moderately & young  \\
     & population  & population  &  young       &        \\
     &  (Gyr)      &  (Gyr)      &  (Myr)       & (Myr) \\
\hline
\multicolumn{5}{l}{ONeMg Novae}\\
LMC 1981     & 10 & 3.2\,--\,1.3 &  - & 125  \\
LMC 1988\#2  & 10 & 3.2\,--\,1.6 & 400 & 79         \\
LMC 1990\#1  & 10 &\multicolumn{2}{c}{4.0 Gyr\,--\,250 Myr} & 40 \\
\\
\multicolumn{5}{l}{Fast} \\
LMC 1977\#2  & 10 & 4.0\,--\,1.0 & 300 & 40  \\
LMC 1987     & 10 & 3.2\,--\,1.3 & 300 & 63 \\
LMC 1992     & - & 2.5\,--\,1.3 & 500 & 63  \\
\\
\multicolumn{5}{l}{Moderately Fast}\\
LMC 1936     & - & 2.0\,--\,1.0 & 300 & 63  \\
LMC 1988\#1  & - & 2.5\,--\,1.0 & 300 & 100 \\
\\
\multicolumn{5}{l}{Slow} \\
LMC 1948     & \multicolumn{3}{c}{10 Gyr\,--\,630 Myr} & 100 \\
\\
\multicolumn{5}{l}{Speed class unknown}\\
LMC 1970\#1  & - & 3.2\,--\,1.0 & 300 & 100   \\
LMC 1973     & 10 & 4.0\,--\,0.80 & - & 63  \\
LMC 1978\#2  & - & 2.5\,--\,1.0 & 400 & 63  \\
LMC 1995     & - & 4.0\,--\,1.3 & 300 & 100  \\
LMC 1997     & 15 & 3.2\,--\,0.80 & - & 100  \\
LMC 2000     & - & 3.2\,--\,0.80 & - & 100 \\
\hline
\end{tabular}
\end{table*} 

Traces of stellar population belonging to the old population of the LMC is 
found in 8 regions. In the remaining 7 regions, no old population is found.  
This indicates a low density of the old population in these regions.  

\begin{figure*}
\centering
\includegraphics[width=17cm]{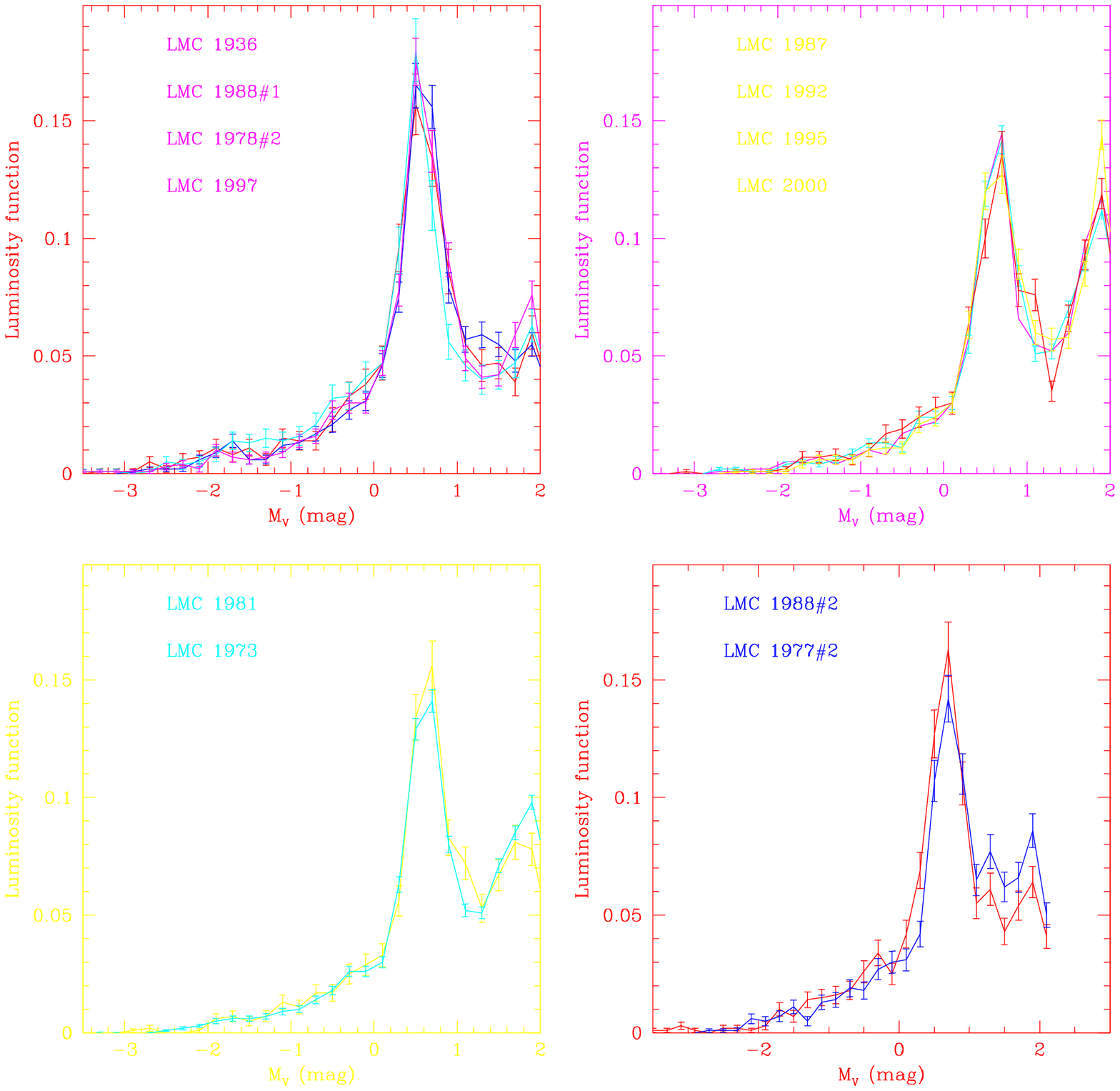}
\caption{The regions having similar profile of RGLF are grouped and plotted. The
error bars indicate the statistical error in the data.}
\label{figure17}
\end{figure*}
It can be seen from table 4 that the intermediate age population seems to 
be in the age range 4.0\,--\,1.0 Gyr, with minor variations depending on 
the region. No detectable population is seen between 4.0\,--\,10.0 Gyr, 
except in one region, around LMC 1948. 

Among the intermediate stellar population, the most prominent feature is the RGC. 
From the data presented in table 3, it is seen that most of the regions have 
around 40\%\,--\,50\% of the total stars in the RGB, with more than 40\% of them in
the RGC.  It is evident from the table that
a higher fraction of RGB stars does not imply a higher fraction of RGC stars.
In can also be noticed that the fraction of RGC stars seem to be similar in the 
case of a some regions, indicative of some groups. 
This could indicate similar star formation events, if the RGLF has similar
profiles. The RGLFs of the regions having similar values for the RGC star 
fraction were
over plotted and some were found to have very similar profiles. 
These are shown in Figure~\ref{figure17}.
The RGLFs are normalised to the total number of RGB stars for comparison.

The top left panel in the figure shows the regions around the novae,
LMC 1936, LMC 1988\#1, LMC 1978\#2 and LMC 1997. The RGLF in these
regions is characterised by a very prominent RGC clump with a narrow and 
peaked profile for the clump. The first two regions in this group are 
located near moderately fast novae and the other two are around novae 
without speed class estimates. The first three regions are located in 
the north-western part of the LMC, whereas the fourth one is located to the 
south-east of the Bar. Also, the first three regions experienced major 
star formation event slightly later, at $\sim$ 2.5 Gyr, which ended by about 
1 Gyr. A few studies on the stellar population in the northern region of
the LMC also show similar results. For e.g.\ the SFH of two northern regions 
studied by Geha et al. (\cite{ge98}) using HST data, showed an enhancement 
in star formation at about 2 Gyr ago. Similarly, the SFH of two other 
northern LMC fields studied by Dolphin (\cite{d00}) provide evidence for an 
increment in star formation about 2.5 Gyr ago.

The top right panel in figure~\ref{figure17} shows that the RGLF of stars 
in the regions around the 
novae, LMC 1987, LMC 1992, LMC 1995 and LMC 2000. These profiles are
characterised by a comparatively broader RGC clump profile, with a shorter 
peak. Among these four regions, the first two are located around fast novae,
whereas the speed class is not known for the novae in the other two regions. 
It is of interest to note that all these four regions are located close to 
the center of the Bar, at least in the projected view. Especially, the 
regions around novae LMC 1987, LMC 1995 and LMC 2000 are located very
close to one another.

The two plots in the upper portion of the figure~\ref{figure17} seem to 
indicate that
there may be a difference in the RGLF of the regions surrounding fast novae 
and moderately fast novae. This is only an indication as the known number 
of fast and moderately fast novae is quite small.

The bottom left panel in figure~\ref{figure17} shows that the RGLF of 
regions around LMC 1981 and LMC 1973 are similar. These are characterised 
by a slightly flatter peak. The first region is around a fast nova, while
the speed class of the nova in the second region is not estimated.
The bottom right panel shows the RGLF of regions around LMC 1988\#2 and 
1977\#2. The RGLF is characterised by a relatively broad and sharp RGC 
clump. These two regions are located around fast novae. 

\begin{figure}
\resizebox{\hsize}{!}{\includegraphics{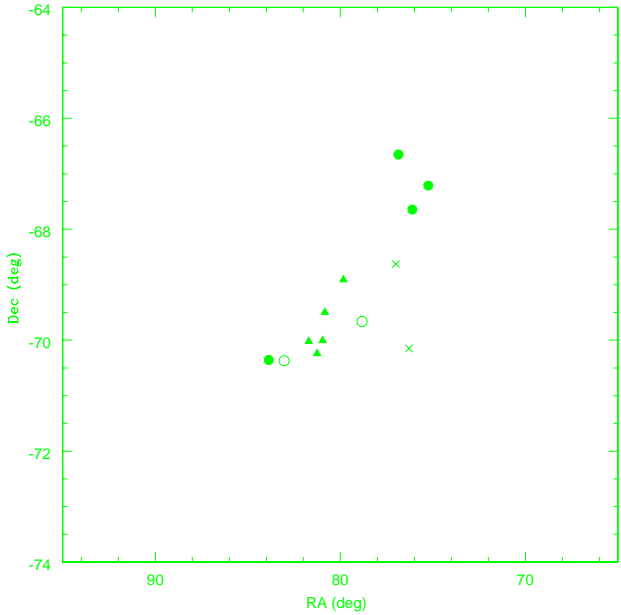}}
\caption{The location of regions considered in figure 17 are plotted here.
 Each group of regions with similar RGLF profiles is identified by the same symbol.}
\label{figure18}
\end{figure}

In order to substantiate the results of the above discussion, the locations of
the groups showing similar RGLF profiles are shown in Figure~\ref{figure18}.
It can be seen from the figure that the novae regions near the central 
region of the Bar have similar RGLF profiles, indicating that these regions 
might have had a similar SFH. A look at the MSLF of these five regions 
indicate that the MSLF profiles are also very similar supporting the fact 
that these regions might have had a similar SFH. 

A striking feature noticed in the analyses is that the stellar population 
near LMC 1948 is different from those seen in other regions. The 
region around LMC 1948 seems to indicate the presence of old stellar 
population, much older than 4.0 Gyr, unlike what is seen in other regions. 
A careful look at the CMDs presented by Udalski et al. (\cite{u2000}) of 
the LMC fields indicate the presence of regions which have a SFH very similar
to that around nova LMC 1948. The HST data of the field region near the 
star cluster NGC 2019, which is located within the Bar, presented by Olsen 
(\cite{Ol99}), also shows a very broad RGC and scattered RGB similar to the
region around nova LMC 1948. This region was also estimated to have had an
enhanced star formation 5-8 Gyr ago. It is thus quite likely that pockets of
regions with a SFH similar to that around LMC 1948 can be found within or 
near the Bar. 

The moderately young population of a few hundred Myr is seen in 10 out of 
15 regions studied here. This is in the age range 250\,--\,500 Myr, with 
6 out of the 10 regions having 300 Myr age population. The youngest stellar 
population is in the range 50\,--\,100 Myr, except in the case of three 
regions. None of the regions show stars younger than 40 Myr.

The metallicity of the isochrones used is Z=0.008, which is equivalent to [Fe/H] = $-$0.40.
This value of metalicity is found to fit all the 15 regions. Hence we find that the
metalicity of the regions studied is more or less the same for the intermediate age stars 
and for stars younger than that.

\paragraph {Red Clump Stars}
The red clump stars correspond to
the core-helium burning stars. When there is a heterogeneous population
of stars, the luminosity of the red clump stars is dependent on the age and
metallicity of the underlying stellar population (Girardi \& Salaris \cite{gs01}).
Hence the age of the RGC stars can be estimated from their luminosity, once the
metallicity is known. Therefore the peak luminosity in the RGC profile can be
used to estimate the approximate age of the RGC stars. As the $Z= 0.008$ is 
found to fit the CMDs, this value of $Z$ can be assumed for the red clump 
stars also. The relation between the age and the luminosity of the red clump 
stars for $Z= 0.008$ given in Girardi \& Salaris (\cite{gs01}) is used to 
estimate the approximate age of the red clump stars in the regions studied 
here. The value of $Z$ assumed here is consistent with the
[Fe/H] values used by Girardi \& Salaris (\cite{gs01}) for the LMC regions. 
It can be seen that the peak luminosity of the RGC profile is in the range
$M_V$ = 0.5\,--\,0.7 mag. The RGLF indicates that either the RGC peaks at
both these magnitudes or at one of these. From table 1 in Girardi \& Salaris
(\cite{gs01}), it can be seen that this range in luminosity corresponds to an 
age range  3\,--\,1 Gyr approximately. Girardi \& Salaris (\cite{gs01}) mention
that the bulk of the red giants in the LMC are $\le$ 3 Gyr. Therefore the red 
clump age range estimated here is within the expected range.

\paragraph {Star clusters}
Star clusters within a radius of 30 arcmin from the nova location are studied
here with the assumption that star cluster formation episodes could be 
linked to star formation 
episodes. Most of the clusters for which age estimates are available appear 
to have formed between 100 Myr and 300 Myr or between 300 Myr and 1 Gyr. 
Only in one case, in the vicinity of LMC 1990\#1, a significant number of 
star clusters appear to have formed before 1 Gyr. 

The star cluster data presented in table 2 shows that though most of the regions
have star clusters, the ages are known for a significant fraction in only some 
regions.  Considering only regions for which ages are known for good fraction of
clusters, we find that 7 out of 13 regions (53.9\%) have clusters aged more than
1 Gyr. As the LMC is seen to be
devoid of star clusters within the age range 3 Gyr to 10 Gyr, and only a few 
very old clusters are known, it is assumed that these clusters fall in the 
1 Gyr\,--\,2.5 Gyr age range. 
The intermediate age star cluster formation episodes are correlated with 
star formation episodes in only $\sim$ 54\% of the regions. It thus appears 
that star clusters are not good tracers of the intermediate age stellar 
population in the LMC.

The star cluster formation seems to be very well correlated to the star 
formation events in the 100\,--\,300 Myr range. A good number of star clusters 
are found in this age range, wherever there has been a noticeable stellar 
population of the same age range. Therefore, the star clusters are seen to be 
good tracers of star formation in the age range 100\,--\,300 Myr.

The youngest cluster population seen is younger than 30 Myr, and is found in 
the vicinity of 7 novae. On the other hand, no stellar population younger than 
40 Myr is seen in any of the novae regions. Therefore, we do not see any 
correlation between cluster formation and star formation in this age range. 
Hence the star clusters are not a good tracers of stellar population younger 
than 30 Myr.

\subsection{Probable parent population of novae}
The parent population of the novae is believed to belong to the intermediate
age population. In the nova regions studied here, we find that the
star formation began a few Gyr ago, though there are traces of old stellar 
population. The population which is the outcome of this
star formation event is likely to be the parent population of the novae.
The star clusters do not appear to be good tracers of the stellar population in 
the intermediate age range, as seen in the previous section. Hence only the
intermediate age stellar population and their ages are considered to 
identify the probable parent population of the LMC novae.
The stellar population considered is the projected population near the
novae. As the LMC is known to have a thin disc, we assume that the depth of the LMC
in the line of sight to be small. Therefore the assumption that the nova and 
the projected stellar population are not verymuch spatially separated is valid.
In this study, we have analysed the stellar population around 15 regions on the
face of the LMC and identify the intermediate population which is common to the
regions belonging to each type of nova. This approach thus more or less
eliminates any biases arising due to the projection effect.

Of the six fast nova regions studied here, we find that in three regions, the intermediate
age population has an upper age limit of 3.2 Gyr, two regions have 4.0 Gyr and
one has 2.5 Gyr. The lower limit is 1.3 Gyr for three, 1.6 Gyr for one and 1 Gyr for
one. As one region (LMC 1990\#1) has continued star formation, the lower age limit
is not possible to identify. This implies that the parent population of fast
novae lie in the age range 4.0\,--\,1 Gyr. Assuming the parent population 
to be in the age range which is in common in all the six regions, the age range 
of the parent population of the fast novae is likely to be 3.2\,--\,1.6 Gyr. 

The regions around the two moderately fast novae studied here indicate a
parent population in the age range 2.5\,--\,1.0 Gyr.

The region around only one slow novae has been studied. The striking differences 
between this region and those around the fast and moderately fast novae are:
(a) continuous star formation between 10 Gyr\,--\,630 Myr, and (b) presence of 
stars belonging to the age range 4\,--\,10 Gyr. As there is only one slow 
nova in the LMC for which field star data are available, we looked into the 
projected stellar population around a slow nova in Small Magellanic Cloud 
(SMC); SMC 1994. SMC 1994 was detected in the EROS microlensing survey,
located in the core of SMC (de Laverny et al \cite{del98}). Detailed
observations of this nova indicated it to be a slow nova, with properties 
similar to the slow, dust forming Galactic novae. The region in which
SMC 1994 occurred has been scanned by the OGLE project in their survey of the 
SMC (Udalsky et al. \cite{u98}). An analyses of the CMD of the field stars
in the nova neighbourhood shows that the region had a SFH similar to the
LMC 1948 region and that the RGB stars span a range of 1\,--\,10 Gyr in age. 
The above facts imply that the progenitors of slow novae probably belong to 
an  older population, consistent with the idea of slower, fainter, bulge 
novae in the Galaxy and M31.

The age estimates for the novae for which the speed class is not known, 
show that the common age range is 1.3\,--\,2.5 Gyr. If one includes all the 
ranges, then the age of the intermediate age population is in the range 
4 Gyr\,--\,800 Myr.

Combining the results obtained for 14 regions around novae, (excluding the 
region around slow nova), we see from table 4, that 28.6\% (4 out of
14 regions) of the regions have 4 Gyr, 42.9\% (6 out of 14) have 3.2 Gyr,
21.4\% (3 out of 14) have 2.5 Gyr and 7\% have 2.0 Gyr as the upper limit 
for age of the intermediate age population. The lower limit for the age
of the intermediate age population is found to be 1.6 Gyr for 7\% of the regions,
1.3 Gyr for 28.6\% of the regions, 1.0 Gyr for 42.9\% of the regions and
800 Myr for 21.4\% of the regions. The important limit derived in this study is 
the upper age limit of the parent population of novae in the LMC, which is 4 Gyr. 
The lower age limit for the parent population
of the novae is more likely to be 1 Gyr. 
If we consider the most likely limit, then the progenitor age range is likely
to be between 3.2\,--\,1.0 Gyr.
The progenitor for the slow novae is likely to originate from the population,
in the age range 1\,--\,10 Gyr.

In the previous section, it was found that the red clump stars in the LMC are
likely to be in the age range 3\,--\,1 Gyr and this is very similar to the 
age range for the progenitors of the fast and moderately fast novae. It is 
very likely that the progenitors of fast and moderately fast novae and red 
clump stars are similar in age and metallicity.

Applying standard models for the formation and evolution of cataclysmic variables,
Kolb \& Stehle (\cite{ks96}) determined the age structure for a model population of 
Galactic cataclysmic variables. Their model predicts that there are two age 
groups, above and below the orbital period gap. The systems above the period gap 
are younger than 1.5 Gyr and those below the gap are of age 3\,--\,4 Gyr. The 
possible age for the nova population derived in this study is in good agreement
with the age estimates of Kolb \& Stehle (\cite{ks96}), keeping in mind that their
estimates are for the Galactic cataclysmic variables. This indicates that 
the progenitor age range as found in the LMC is very 
similar to that estimated for our Galaxy. As the novae in the LMC are 
predominantly disk novae, the progenitors of disk novae are likely
to be in the age range 3.2\,--\,1 Gyr.

\paragraph{Other galaxies} The SFH around the nova regions in the LMC may be 
compared with the SFH of other galaxies where novae have been detected, in
particular, M33 which is also a disk dominated galaxy. Based on a study
of the star clusters in M33, Chandar et al. (\cite{ch99}) and Ma et al.
(\cite{ma01}) find that M33 experienced an enhancement in cluster formation 
around a few Gyr and that most of these clusters are metal poor. If we assume 
that the intermediate age cluster formation is correlated with that of the
stellar population, then it appears that the parent population of novae in
M33 are similar to those in the LMC. It is of further interest to note that 
NGC 5128, an elliptical galaxy where a fair number of novae have been
detected (Ciardullo et al. \cite{cia90}), shows a significant presence of
intermediate age AGB stars as indicated by a recent study based on VLT
observations (Rejkuba et al. \cite{rej01}).

\section{Summary and conclusions}

The underlying population in the location of novae in the LMC are studied
in an attempt to understand the common characteristics of the nova regions and
thereby tried to identify the parent population of novae. There has been no
previous attempt to directly study the stellar population in the neighbourhood
of novae. 

The LMC is an ideal galaxy to undertake such a study as a good number of novae are 
identified in the LMC and the field population has been very well studied through 
many recent surveys.
The stellar population and the star clusters near the novae are used for this study.
The local population of the novae are studied using the OGLE II field star data and
the star clusters are obtained from catalogues. The star formation history in the
location of 15 novae are studied using CMDs, LFs and comparing the fraction of 
stars in the MS, RGB and RGC. The episodes of cluster formation are determined by 
grouping the clusters in age. The above analysis leads to the following
conclusion:
\begin{enumerate}
\item The LMC has 72.2\% novae in the fast category and 11.1\% in the slow category.
This suggests that the fraction of the intermediate age stellar population to 
which the fast novae belong is correspondingly substantial, when
compared to the fraction of the old stellar population which are responsible 
for the slow novae.
\item The fast novae (including the ONeMg type) and the moderately fast novae 
have a very similar parent population. Also, the SFH of intermediate age stars
near these novae are very similar, while that around the slow nova is very
different. 
\item
The upper age limit of the stellar population
to which the fast and moderately fast novae belong is likely to be 3.2 Gyr, 
while it cannot be higher than 4 Gyr.
The lower age limit of the parent nova population is most likely to be
1.0 Gyr. This is in good agreement with the age estimates of Kolb 
\& Stehle (\cite{ks96}) for Galactic cataclysmic variables.
\item The region around slow nova shows continuous star 
formation between 1\,--\, 10 Gyr, with a good fraction belonging to the 
4\,--\,10 Gyr population, consistent with the idea that the progenitors of
slow novae belong to an older population.
\end{enumerate}

\begin{acknowledgements}
We thank R. Sagar, H.W. Duerbeck and the referee for useful comments and suggestions.
\end{acknowledgements}

\end{document}